\pdfoutput=1
\documentclass[acmsmall]{acmart}
\usepackage{booktabs}
\usepackage[ruled,vlined,linesnumbered]{algorithm2e}
\usepackage{dashbox}
\usepackage{calc}
\usepackage{tikz}
\usetikzlibrary{calc}
\usetikzlibrary{decorations.markings}
\usetikzlibrary{decorations.pathreplacing}
\usetikzlibrary{intersections}
\usetikzlibrary{patterns}
\usetikzlibrary{shapes.misc}
\usepackage{enumitem}
\setlist{leftmargin=2em}

\makeatletter
\fancypagestyle{standardpagestyle}{
  \fancyhf{}
  \fancyhead[LE]{\ACM@linecountL\thepage}
  \fancyhead[RO]{\thepage}
  \fancyhead[RE]{\@shortauthors}
  \fancyhead[LO]{\ACM@linecountL\shorttitle}
}
\fancypagestyle{firstpagestyle}{
  \fancyhf{}
  \fancyhead[L]{\ACM@linecountL}
}
\def\maketitle{
  \@topnum\z@ % this prevents figures from falling at the top of page 1
  \@botnum\z@ % we do not want them to be on the bottom either
  \begingroup
  \let\@footnotemark\@footnotemark@nolink
  \let\@footnotetext\@footnotetext@nolink
  \renewcommand\thefootnote{\@fnsymbol\c@footnote}
  \hsize=\textwidth
  \def\@makefnmark{\hbox{\@textsuperscript{\@thefnmark}}}
  \@mktitle\@mkauthors\@mkteasers
  \@printtopmatter
  \setcounter{footnote}{0}
  \def\@makefnmark{\hbox{\@textsuperscript{\normalfont\@thefnmark}}}
  \@titlenotes\@subtitlenotes\@authornotes
  \let\@makefnmark\relax\let\@thefnmark\relax
  \let\@makefntext\noindent
  \ifx\@empty\thankses\else\footnotetextcopyrightpermission{\def\par{\let\par\@par}\parindent\z@\@setthanks}\fi
  \endgroup
  \setcounter{footnote}{0}
  \@mkabstract
  \andify\authors
  \andify\shortauthors
  \global\let\authors=\authors
  \global\let\shortauthors=\shortauthors
  \hypersetup{pdfauthor={\authors},pdftitle={\shorttitle}}
  \@printendtopmatter
  \@afterindentfalse
  \@afterheading
}
\makeatother

\tikzset{
  on each segment/.style={decorate,decoration={
    show path construction,
    moveto code={},
    lineto code={\path[#1](\tikzinputsegmentfirst)--(\tikzinputsegmentlast);},
    closepath code={\path[#1](\tikzinputsegmentfirst)--(\tikzinputsegmentlast);}
  }},
  mid arrow/.style={postaction={decorate,decoration={
    markings,mark=at position .5 with {\arrow[#1]{stealth}}
  }}}
}

\newcommand{\explanatorybox}[2]{\smash{\dbox{\parbox[t]{#1}{#2}}}}

\newcommand{\setR}{\mathbb R}
\newcommand{\setZ}{\mathbb Z}
\newcommand{\HHH}{\mathcal H}
\newcommand{\LLL}{\mathcal L}
\newcommand{\PPP}{\mathcal P}
\newcommand{\ZZZ}{\mathcal Z}
\newcommand{\TRUE}{\mathtt{true}}
\newcommand{\FALSE}{\mathtt{false}}
\newcommand{\dets}{\det\nolimits^\star}

\DeclareMathOperator{\sgn}{sgn}
\DeclareMathOperator{\LHP}{LHP}
\DeclareMathOperator{\RHP}{RHP}
\DeclareMathOperator{\wideLHP}{\makebox[\widthof{$\RHP$}][c]{$\LHP$}}

\let\setminus\smallsetminus
\let\Delta\varDelta
\let\Theta\varTheta
\let\Omega\varOmega

\AtEndPreamble{\newtheorem{observation}[theorem]{Observation}}

\begin{document}

\title[Common tangents of two disjoint polygons in linear time and constant workspace]{Common Tangents of Two Disjoint Polygons in Linear Time and Constant Workspace}
\author{Mikkel Abrahamsen}
\affiliation{\institution{University of Copenhagen}\country{Denmark}}
\email{miab@di.ku.dk}
\author{Bartosz Walczak}
\affiliation{\institution{Jagiellonian University}\country{Poland}}
\email{walczak@tcs.uj.edu.pl}

\begin{abstract}
We provide a remarkably simple algorithm to compute all (at most four) common tangents of two disjoint simple polygons.
Given each polygon as a read-only array of its corners in cyclic order, the algorithm runs in linear time and constant workspace and is the first to achieve the two complexity bounds simultaneously.
The set of common tangents provides basic information about the convex hulls of the polygons---whether they are nested, overlapping, or disjoint---and our algorithm thus also decides this relationship.
\end{abstract}

\thanks{%
  A journal version of this paper appeared in \href{https://doi.org/10.1145/3284355}{\textit{ACM Trans.\ Algorithms} 15, 1 (2018), 12:1--12:21, doi:10.1145/3284355}.

  Preliminary expositions of the results contained in this paper appeared at SoCG 2015 \cite{Abr15} and ESA 2016 \cite{AW16}.

  Mikkel Abrahamsen was partially supported by Danish Council for Independent Research grant DFF-0602-02499B.

  Bartosz Walczak was partially supported by National Science Center of Poland grant 2015/17/D/ST1/00585.

  Authors' addresses:
  Mikkel Abrahamsen, Department of Computer Science, University of Copenhagen, Denmark, e-mail:\ \href{mailto:miab@di.ku.dk}{\texttt{miab@di.ku.dk}};
  Bartosz Walczak, Department of Theoretical Computer Science, Faculty of Mathematics and Computer Science, Jagiellonian University, Krak\'ow, Poland, e-mail:\ \href{mailto:walczak@tcs.uj.edu.pl}{\texttt{walczak@tcs.uj.edu.pl}}.
}

\maketitle

\section{Introduction}

A tangent of a polygon is a line touching the polygon such that all of the polygon lies on the same side of the line.
We consider the problem of computing the common tangents of two disjoint polygons that are simple, that is, they have no self-intersections.
The set of common tangents provides basic information about the convex hulls of the polygons---whether they are disjoint, overlapping, or nested.
We call a common tangent \emph{outer} if the two polygons lie on the same side of it and \emph{separating} otherwise.
Two disjoint polygons have two outer common tangents unless their convex hulls are nested, and if they are properly nested, then there is no outer common tangent.
Two polygons have two separating common tangents unless their convex hulls overlap, and if they properly overlap, then there is no separating common tangent.
See Figures \ref{fig:disjoint}--\ref{fig:nested} for illustrations.
Common tangents arise in many different contexts, for instance in problems related to convex hulls \cite{PH77}, shortest paths \cite{GH89}, ray shooting \cite{HS95}, and clustering \cite{AdBB+17}.

\begin{figure}[t]
\begin{minipage}[t]{2.1in}
\centering
\begin{tikzpicture}[line cap=round,line join=round,scale=0.27,yscale=1.1]
  \draw (2.54,8.12)--(2.84,6.22)--(4.45,5.5)--(4.02,6.6)--(5.26,9.04)--(5.34,6.26)
    --(6.46,5.26)--(2.26,3.52)--(6.46,3.14)--(2.52,2.44)--(4.44,2.08)--(1.16,1.24)
    --(0.2,4.02)--(3.68,4.74)--(1.68,5.04)--(0.52,7.74)--(1.7,7.48)--cycle;
  \draw[blue] (10.16,9.26)--(8,7.6)--(8.54,5.32)--(10.56,6.96)--(12.5,6.7)--(12.12,2.96)
    --(9.66,5.1)--(8.48,2.32)--(11,1.24)--(14.54,2.14)--(16.04,4.9)--(13.5,2.94)
    --(13.6,7.12)--(16,6.88)--(14.8,8.5)--(11.22,7.72)--(12.64,8.7)--cycle;
  \path (5.26,9.04)--(10.16,9.26) coordinate[pos=-1.1] (A) coordinate[pos=2.2] (B);
  \path (1.16,1.24)--(11,1.24) coordinate[pos=-0.1] (C) coordinate[pos=1.45] (D);
  \path (5.26,9.04)--(8.48,2.32) coordinate[pos=-0.15] (E) coordinate[pos=1.3] (F);
  \path (6.46,3.14)--(8,7.6) coordinate[pos=-0.6] (G) coordinate[pos=1.6] (H);
  \draw[dashed] (A)--(B);
  \draw[dashed] (C)--(D);
  \draw[dashed] (E)--(F);
  \draw[dashed] (G)--(H);
\end{tikzpicture}
\caption{The convex hulls are disjoint---outer and separating common tangents exist.\\\strut}
\label{fig:disjoint}
\end{minipage}\hfill
\begin{minipage}[t]{1.55in}
\centering
\begin{tikzpicture}[line cap=round,line join=round,scale=1.2]
  \draw (-5:0.8mm)\foreach\i in {2,...,25}{--(\i*30:\i*0.5mm)}--
    (775:13.5mm)\foreach\i in {25,...,1}{--(\i*30:\i*0.5mm+1mm)}--cycle;
  \draw[blue] (175:0.8mm)\foreach\i in {2,...,22}{--(\i*30+180:\i*0.5mm)}--
    (865:12mm)\foreach\i in {22,...,1}{--(\i*30+180:\i*0.5mm+1mm)}--cycle;
  \path (775:13.5mm)--(810:11.5mm) coordinate[pos=-0.7] (A) coordinate[pos=2.8] (B);
  \draw[dashed] (A)--(B);
  \path (865:12mm)--(540:10mm) coordinate[pos=-1.1] (C) coordinate[pos=2.7] (D);
  \draw[dashed] (C)--(D);
\end{tikzpicture}
\caption{The convex hulls over\-lap---only outer common tangents exist.\strut}
\end{minipage}\hfill
\begin{minipage}[t]{1.3in}
\centering
\begin{tikzpicture}[line join=round,scale=0.3]
  \draw (-1.74,1.74)--(0.18,2.06)--(-0.78,-1.4)--(-1.06,0.22)--(-2.4,-0.9)
    --(-0.88,-3.4)--(4.48,-2.3)--(6.74,2.92)--(3.64,-1.44)--(1,-1.06)
    --(1.16,1.92)--(-0.18,4.28)--cycle;
  \draw[blue] (3.96,0.52)--(3.8,-0.36)--(2.42,0.12)--(2.7,1.86)--(2.14,2.62)
    --(2.24,1.12)--(1.04,3.56)--(3.62,2.62)--(3.14,0.82)--(5.24,2.02)--cycle;
\end{tikzpicture}
\caption{The convex hulls are nested---no common tangents exist.\strut}
\label{fig:nested}
\end{minipage}
\end{figure}

We provide a very simple algorithm to compute the common tangents of two disjoint simple polygons.
In view of the above, the algorithm also determines whether the two polygons have (properly) nested, (properly) overlapping, or disjoint convex hulls.
Given each of the two polygons as a read-only array of its corners in cyclic order, our algorithm runs in linear time and uses seven variables each storing a boolean value or an index of a corner in one of the arrays.
The algorithm is therefore asymptotically optimal with respect to time and workspace, and it operates in the \emph{constant workspace model} of computation.

The constant workspace model is a restricted version of the RAM model in which the input is read-only, the output is write-only, and only $O(\log b)$ additional bits of \emph{workspace} (with both read and write access) are available, where $b$ denotes the bit length of the input.
It is natural to consider algorithms in this model as memory-optimal, because $\Omega(\log b)$ bits are required to store a pointer to an entry in the input.
Since blocks of $\Theta(\log b)$ bits are considered to form \emph{words} in the memory, algorithms in the constant workspace model use $O(1)$ words of workspace, which explains the name of the model.
(Likewise, time complexity is usually measured with respect to the number of words in the input, with the assumption that arithmetic operations on words can be performed in constant time.)
The practical relevance of studying problems in the constant workspace model is increasing, as there are many current and emerging memory technologies where writing can be much more expensive than reading in terms of time and energy \cite{CDG+16}.

The constant workspace model was first considered explicitly for geometric problems by \citeN{AMRW11}.
Recently, there has been growing interest in algorithms for geometric problems using constant or restricted workspace.
We refer the reader to the recent survey by \citeN{BKM18} for an overview of the results.
In complexity theory, the class of decision problems solvable using constant workspace is usually denoted by $\mathsf{L}$.
The constant workspace model was shown to be surprisingly powerful---for instance, the problem of deciding whether two vertices of an undirected graph lie in the same connected component belongs to $\mathsf{L}$~\cite{Rei08}.

The problem of computing common tangents of two polygons has received much attention in the special case that the polygons are convex.
For instance, computing the outer common tangents of disjoint convex polygons is used as a subroutine in the classical divide-and-conquer algorithm for the convex hull of a set of $n$ points in the plane due to \citeN{PH77}.
They gave a naive linear-time algorithm for outer common tangents, which suffices for an $O(n\log n)$-time convex hull algorithm.
The problem is also considered in various dynamic convex hull algorithms \cite{BJ02,HS92,OvL81}.
\citeN{OvL81} gave an $O(\log n)$-time algorithm for computing an outer common tangent of two disjoint convex polygons when a separating line is known, where each polygon has at most $n$ corners.
\citeN{KS95} gave an $O(\log n)$-time algorithm for the same problem but without using a separating line.
\citeN{GHS91} gave a lower bound of $\Omega(\log^2 n)$ on the time required to compute an outer common tangent of two intersecting convex polygons even when they are known to intersect in at most two points.
They also described an algorithm achieving that bound.
\citeN{Tou83} considered the problem of computing separating common tangents of convex polygons.
He gave a linear-time algorithm using the technique of ``rotating calipers''.
\citeN{GHS91} gave an $O(\log n)$-time algorithm for the same problem.
All the above-mentioned algorithms with sublinear running times make essential use of convexity of the polygons.
If the polygons are not convex, a linear-time algorithm can be used to compute the convex hulls before computing the tangents.
Many such algorithms have been described, and the one due to \citeN{Mel87} is usually considered the simplest.
However, if the polygons are given in read-only memory, then $\Omega(n)$ extra bits are required to store the convex hulls, so this approach does not work in the constant workspace model.

In the following, we provide a brief description of our algorithm, which is presented in full detail using the pseudocode in Algorithm~\ref{ALG2} on  page~\pageref{ALG2}.
Algorithm~\ref{ALG1} on page~\pageref{ALG2} is a simplified version of Algorithm~\ref{ALG2}, which finds the separating common tangents whenever they exist, but is guaranteed to find the outer common tangents only when the convex hulls of the two polygons are disjoint.

In order to find a particular common tangent of two polygons $P_0$ and $P_1$, we maintain a pair of corners of support $q_0\in P_0$ and $q_1\in P_1$, where the line $\LLL(q_0,q_1)$ passing through $q_0$ and $q_1$ is considered as a candidate for the requested common tangent.
In each step of the algorithm, we traverse each polygon in order to find a corner that does not lie on the ``correct side'' of the candidate line $\LLL(q_0,q_1)$, that is, on the side where all of the polygon should lie if $\LLL(q_0,q_1)$ was the requested common tangent.
Each polygon is traversed starting from its current corner of support in a direction determined by the type of the tangent that we aim to find---the separating common tangents are computed by traversing both polygons in the same direction, whereas the outer common tangents are computed by choosing opposite directions (see Figure~\ref{fig:assumptions} on page~\pageref{fig:assumptions}).
If both polygons lie entirely on the ``correct side'' of $\LLL(q_0,q_1)$, then we return $\LLL(q_0,q_1)$ as the solution.
Otherwise, when we first encounter an edge $e$ of one of the polygons, say $P_0$, that ends at a corner $q_0'$ on the ``wrong side'', we distinguish two cases, determined by where $e$ intersects the line $\LLL(q_0,q_1)$ with respect to $q_0$ and $q_1$.
If $q_1$ does \emph{not} lie between $q_0$ and the intersection point of $e$ and $\LLL(q_0,q_1)$, then $q_0'$ becomes the new corner of support of $P_0$ for the next step of the algorithm (the corner of support of $P_1$ remains at $q_1$).
Otherwise, $q_1$ lies in the convex hull of $P_0$ and therefore cannot be a support of a common tangent.
In the latter case, we temporarily block $q_0$ from further updates until the first update to $q_1$.
If no update to $q_1$ occurs before a full traversal of $P_1$, we conclude that the convex hulls are nested and no common tangents exist.
The key observation is that if the requested common tangent exists, then it must be found before either polygon has been fully traversed for the second time by its corner of support.
Therefore, if an update occurs during the third full traversal of a polygon, we conclude that the common tangent does not exist.

In order to guarantee a linear bound on the total running time, in each step, the search for a corner on the ``wrong side'' is performed by a tandem walk on the two polygons.
That is, we traverse both polygons starting from the current corners of support and advancing alternately by one edge until finding the first corner on the ``wrong side'' in either of the two polygons---that corner becomes the start of the next search on that polygon (as the new corner of support), while the search on the other polygon is reverted to where it started.
To our knowledge, the idea of a tandem walk was first applied by \citeN{ES81} to a problem not related to geometry.

\citeN{BKL+15} describe a linear-time constant-workspace algorithm, attributed to A.~Pilz, for the following problem: given a simple polygonal chain $P$ with endpoints on its convex hull, and given a line $L$ that separates the two endpoints of $P$, find the two edges of the convex hull of $P$ that are crossed by $L$.
It applies an analogous principle of updating the candidate line by parallel traversal of two independent parts of $P$.
These updates, however, make the points of support of the candidate line move only towards the endpoints of $P$, never coming back to points visited before.
On the other hand, our algorithm sometimes needs to make more than one full traversal of a polygon (but never more than two) in order to find the requested common tangent, which makes its analysis significantly more involved.
An easy adaptation of Pilz's algorithm can be used to find the outer common tangents of two polygons that are separated by a given line.
However, finding such a line seems to be no easier than computing the separating common tangents.

The rest of the paper is organized as follows.
In Section~\ref{sec:terminology}, we introduce the terminology and conventions used throughout the paper and state some well-known properties of the common tangents of two polygons.
In Section~\ref{sec:description}, we describe two algorithms for computing the common tangents if they exist or detecting that they do not exist, where one is a simplified version of the other for special cases indicated before.
Section~\ref{sec:correctness} contains proofs that the algorithms work correctly under the assumption of a crucial lemma, which is then proved in Section~\ref{sec:main-lemma}.
We conclude in Section~\ref{sec:conclusion} by discussing how to avoid the general position assumption of Sections \ref{sec:terminology}--\ref{sec:main-lemma} and by suggesting some related open problems for future research.

\section{Basic terminology and notation}
\label{sec:terminology}

For any two points $a$ and $b$ in the plane, the closed line segment with endpoints $a$ and $b$ is denoted by $ab$.
When $a\neq b$, the line passing through $a$ and $b$ is denoted by $\LLL(a,b)$.
The segment $ab$ and the line $\LLL(a,b)$ are considered \emph{oriented} in the direction from $a$ towards $b$.
A \emph{simple polygon} or just a \emph{polygon} with \emph{corners} $a_0,\ldots,a_{n-1}$, denoted by $\PPP(a_0,\ldots,a_{n-1})$, is a closed curve in the plane composed of $n$ \emph{edges} $a_0a_1,\ldots,a_{n-2}a_{n-1},a_{n-1}a_0$ that have no common points other than the common endpoints of pairs of edges consecutive in that cyclic order.
The polygon $\PPP(a_0,\ldots,a_{n-1})$ is considered \emph{oriented} so that its \emph{forward} traversal visits corners $a_0,\ldots,a_{n-1}$ in this cyclic order.
A \emph{polygonal region} is a closed and bounded region of the plane whose boundary is a polygon.
For any two points $a=(a_x,a_y)$ and $b=(b_x,b_y)$ in $\setR^2$, we let
\begin{equation*}
\det(a,b)=\begin{vmatrix}
  a_x & b_x \\
  a_y & b_y
\end{vmatrix}=a_xb_y-b_xa_y\text{.}
\end{equation*}
For $a_0,\ldots,a_{n-1}\in\setR^2$, we let
\begin{equation*}
\dets(a_0,\ldots,a_{n-1})=\det(a_0,a_1)+\cdots+\det(a_{n-2},a_{n-1})+\det(a_{n-1},a_0)\text{.}
\end{equation*}
In particular, for any three points $a=(a_x,a_y)$, $b=(b_x,b_y)$, and $c=(c_x,c_y)$ in $\setR^2$, we have
\begin{equation*}
\dets(a,b,c)=\begin{vmatrix}
  a_x & b_x \\
  a_y & b_y
\end{vmatrix}+\begin{vmatrix}
  b_x & c_x \\
  b_y & c_y
\end{vmatrix}+\begin{vmatrix}
  c_x & a_x \\
  c_y & a_y
\end{vmatrix}=\begin{vmatrix}
  a_x & b_x & c_x \\
  a_y & b_y & c_y \\
  1 & 1 & 1
\end{vmatrix}\text{.}
\end{equation*}
For two distinct points $a$ and $b$ in the plane, the \emph{left side} and the \emph{right side} of an oriented line $\LLL(a,b)$ are the two closed half-planes $\LHP(a,b)$ and $\RHP(a,b)$, respectively, defined as follows:
\begin{align*}
\wideLHP(a,b)&=\{c\in\setR^2\colon\dets(a,b,c)\geq 0\}\text{,} \\
\RHP(a,b)&=\{c\in\setR^2\colon\dets(a,b,c)\leq 0\}\text{,}
\end{align*}
where $\LHP$/$\RHP$ stands for ``left/right half-plane''.
An oriented polygon $\PPP(a_0,\ldots,a_{n-1})$ is \emph{counterclockwise} when $\dets(a_0,\ldots,a_{n-1})>0$ and \emph{clockwise} when $\dets(a_0,\ldots,a_{n-1})<0$.

We assume for the rest of this paper that $P_0$ and $P_1$ are two disjoint simple polygons with $n_0$ and $n_1$ corners, respectively, each defined by a read-only array of its corners:
\begin{align*}
P_0&=\PPP(p_0[0],\ldots,p_0[n_0-1])\text{,} & P_1&=\PPP(p_1[0],\ldots,p_1[n_1-1])\text{.}
\end{align*}
We make no assumption (yet) on whether $P_0$ and $P_1$ are oriented counterclockwise or clockwise.
We further assume that the corners of $P_0$ and $P_1$ are in general position, that is, $P_0$ and $P_1$ have no corners in common and the combined set of corners $\{p_0[0],\ldots,p_0[n_0-1],p_1[0],\ldots,p_1[n_1-1]\}$ contains no triple of collinear points.
This assumption simplifies the description and the analysis of the algorithm but can be avoided, as we explain in the last section.
We do not assume the polygonal regions bounded by $P_0$ and $P_1$ to be disjoint---they may be nested.
Indices of the corners of each $P_k$ are considered modulo $n_k$, so that $p_k[i]$ and $p_k[j]$ denote the same corner when $i\equiv j\pmod{n_k}$.

A \emph{tangent} of $P_k$ is a line $L$ such that $P_k$ has a common point with $L$ and is contained in one of the two closed half-planes determined by $L$.
A line $L$ is a \emph{common tangent} of $P_0$ and $P_1$ if it is a tangent of both $P_0$ and $P_1$; it is an \emph{outer common tangent} if $P_0$ and $P_1$ lie on the same side of $L$ and a \emph{separating common tangent} otherwise.
The following lemma asserts well-known properties of common tangents of polygons.
See Figures \ref{fig:disjoint}--\ref{fig:nested}.

\begin{lemma}
\label{lem:folklore}
A line is a tangent of a polygon\/ $P$ if and only if it is a tangent of the convex hull of\/ $P$.
Moreover, under the general position assumption, the following holds:
\begin{itemize}
\item $P_0$ and\/ $P_1$ have no common tangents if the convex hulls of\/ $P_0$ and\/ $P_1$ are nested;
\item $P_0$ and\/ $P_1$ have two outer common tangents and no separating common tangents if the convex hulls of\/ $P_0$ and\/ $P_1$ properly overlap;
\item $P_0$ and\/ $P_1$ have two outer common tangents and two separating common tangents if the convex hulls of\/ $P_0$ and\/ $P_1$ are disjoint.
\end{itemize}
\end{lemma}

\section{Algorithms}
\label{sec:description}

We distinguish four cases of the common tangent problem: find the pair of indices $(s_0,s_1)$ such that
\begin{enumerate}
\item $P_0\subset\RHP(p_0[s_0],p_1[s_1])$ and $P_1\subset\RHP(p_0[s_0],p_1[s_1])$,
\item $P_0\subset\wideLHP(p_0[s_0],p_1[s_1])$ and $P_1\subset\wideLHP(p_0[s_0],p_1[s_1])$,
\item $P_0\subset\RHP(p_0[s_0],p_1[s_1])$ and $P_1\subset\wideLHP(p_0[s_0],p_1[s_1])$,
\item $P_0\subset\wideLHP(p_0[s_0],p_1[s_1])$ and $P_1\subset\RHP(p_0[s_0],p_1[s_1])$.
\end{enumerate}
The line $\LLL(p_0[s_0],p_1[s_1])$ is an outer common tangent in cases 1--2 and a separating common tangent in cases 3--4.
We say that $(s_0,s_1)$ is the \emph{solution} to the particular case of the problem.
An algorithm solving each case is expected to find the solution $(s_0,s_1)$ if it exists (i.e.\ the convex hulls of $P_0$ and $P_1$ are not nested in cases 1--2 and are disjoint in cases 3--4) and to report ``no solution'' otherwise.

We will describe two general algorithms.
Algorithm~\ref{ALG1}, very simple, fully solves the separating common tangent problem (cases 3--4), finding the separating common tangent if the convex hulls of $P_0$ and $P_1$ are disjoint and otherwise reporting that the requested tangent does not exist.
Furthermore, Algorithm~\ref{ALG1} solves the outer common tangent problem (cases 1--2) provided that the convex hulls of $P_0$ and $P_1$ are disjoint.
Algorithm~\ref{ALG1} also correctly reports that the outer common tangents do not exist if the convex hulls of $P_0$ and $P_1$ are nested.
However, Algorithm~\ref{ALG1} can fail to find the outer common tangents if the convex hulls of $P_0$ and $P_1$ properly overlap.
Algorithm~\ref{ALG2} is an improved version of Algorithm~\ref{ALG1} that solves the problem correctly in all cases.

The general idea behind either algorithm is as follows.
The algorithm maintains a pair of indices $(s_0,s_1)$ called the \emph{candidate solution}, which determines the line $\LLL(p_0[s_0],p_1[s_1])$ called the \emph{candidate line}.
If each of the two polygons lies on the ``correct side'' of the candidate line, which is either $\RHP(p_0[s_0],p_1[s_1])$ or $\LHP(p_0[s_0],p_1[s_1])$ depending on the particular case of 1--4 to be solved, then the algorithm returns $(s_0,s_1)$ as the requested solution.
Otherwise, for some $u\in\{0,1\}$, the algorithm finds an index $v_u$ such that $p_u[v_u]$ lies on the ``wrong side'' of the candidate line, updates $s_u$ by setting $s_u\gets v_u$, and repeats.
This general scheme guarantees that if $(s_0,s_1)$ is claimed to be the solution, then it indeed is.
However, the algorithm can fall in an infinite loop---when there is no solution or when the existing solution keeps being missed.
A detailed implementation of the scheme must guarantee that the solution is found in linearly many steps if it exists.
Then, if the solution is not found in the guaranteed number of steps, the algorithm terminates and reports ``no solution''.

\begin{algorithm}[p]
\LinesNumbered
\DontPrintSemicolon
\SetArgSty{}
\SetKwIF{If}{ElseIf}{Else}{if}{}{else if}{else}{end if}
\SetKwFor{While}{while}{}{end while}
$s_0\gets 0$;\quad $v_0\gets 0$;\quad $s_1\gets 0$;\quad $v_1\gets 0$;\quad $u\gets 0$\;
\While{$s_0<2n_0$ and $s_1<2n_1$ and ($v_0<s_0+n_0$ or $v_1<s_1+n_1$)} {
  $v_u\gets v_u+1$\;
  \If {$p_u[v_u]\notin\HHH_u(p_0[s_0],p_1[s_1])$\hfill\explanatorybox{185pt}{
    See Figure~\ref{fig:assumptions} for the meaning of $\HHH_u(a,b)$ and the assumed orientations of $P_0$ and $P_1$.
  }} {\nllabel{line1:side}
    $s_u\gets v_u$;\quad $v_{1-u}\gets s_{1-u}$\;\nllabel{line1:update}
  }
  $u\gets 1-u$\;
}
\If {$s_0\geq 2n_0$ or $s_1\geq 2n_1$} {
  \Return {``no solution''}\;
}
\Return {$(s_0,s_1)$}\;
\caption{}
\label{ALG1}
\end{algorithm}

\begin{algorithm}[p]
\LinesNumbered
\DontPrintSemicolon
\SetArgSty{}
\SetKwIF{If}{ElseIf}{Else}{if}{}{else if}{else}{end if}
\SetKwFor{While}{while}{}{end while}
$s_0\gets 0$;\quad $v_0\gets 0$;\quad $b_0\gets\FALSE$;\quad $s_1\gets 0$;\quad $v_1\gets 0$;\quad $b_1\gets\FALSE$;\quad $u\gets 0$\;
\While{$s_0<2n_0$ and $s_1<2n_1$ and ($v_0<s_0+n_0$ or $v_1<s_1+n_1$)} {
  $v_u\gets v_u+1$\;
  \If {$p_u[v_u]\notin\HHH_u(p_0[s_0],p_1[s_1])$ and not $b_u$\hfill\explanatorybox{128pt}{
    See Figure~\ref{fig:assumptions} for the meaning of $\HHH_u(a,b)$ and the assumed orientations of $P_0$ and $P_1$;
    $\Delta(a,b,c)$ is the triangle spanned by $a$, $b$, $c$.
  }} {\nllabel{line2:side}
    \uIf {$p_{1-u}[s_{1-u}]\in\Delta(p_u[s_u],p_u[v_u-1],p_u[v_u])$} {\nllabel{line2:triangle}
      $b_u\gets\TRUE$\;\nllabel{line2:setb}
    }\Else {
      $s_u\gets v_u$;\quad $v_{1-u}\gets s_{1-u}$;\quad $b_{1-u}\gets\FALSE$\;\nllabel{line2:update}
    }
  }
  $u\gets 1-u$\;
}
\If {$s_0\geq 2n_0$ or $s_1\geq 2n_1$ or $b_0$ or $b_1$} {
  \Return {``no solution''}\;
}
\Return {$(s_0,s_1)$}\;
\caption{}
\label{ALG2}
\end{algorithm}

\begin{figure}[p]
\centering
\begin{minipage}[t]{2.1in}
\centering
\begin{tikzpicture}[line cap=round,line join=round,scale=0.3,yscale=1.1]
  \draw (2.54,8.12)--(2.84,6.22)--(4.45,5.5)--(4.02,6.6)--(5.26,9.04)--(5.34,6.26)
    --(6.46,5.26)--(2.26,3.52)--(6.46,3.14)--(2.52,2.44)--(4.44,2.08)--(1.16,1.24)
    --(0.2,4.02)--(3.68,4.74)--(1.68,5.04)--(0.52,7.74)--(1.7,7.48)--cycle;
  \draw[blue] (10.16,9.26)--(8,7.6)--(8.54,5.32)--(10.56,6.96)--(12.5,6.7)--(12.12,2.96)
    --(9.66,5.1)--(8.48,2.32)--(11,1.24)--(14.54,2.14)--(16.04,4.9)--(13.5,2.94)
    --(13.6,7.12)--(16,6.88)--(14.8,8.5)--(11.22,7.72)--(12.64,8.7)--cycle;
  \path (5.26,9.04)--(10.16,9.26) coordinate[pos=-1.1] (A) coordinate[pos=2.16] (B) coordinate[pos=0.32] (L1);
  \path (1.16,1.24)--(11,1.24) coordinate[pos=-0.1] (C) coordinate[pos=1.45] (D) coordinate[pos=0.63] (L2);
  \path (5.26,9.04)--(8.48,2.32) coordinate[pos=-0.15] (E) coordinate[pos=1.3] (F);
  \path (6.46,3.14)--(8,7.6) coordinate[pos=-0.63] (G) coordinate[pos=1.62] (H);
  \draw[dashed] (A)--(B);
  \draw[dashed] (C)--(D);
  \draw[dashed] (E)--(F);
  \draw[dashed] (G)--(H);
  \path (1.68,5.04)--(0.52,7.74) node[pos=0.2,left] {$P_0$};
  \path (13.5,2.94)--(13.6,7.12) node[pos=0.6,blue,right] {$P_1$};
  \node[above] at (L1) {1};
  \node[below] at (L2) {2};
  \node[above] at (E) {3};
  \node[above] at (H) {4};
\end{tikzpicture}
\end{minipage}\hskip 1in
\begin{minipage}[t]{1.55in}
\centering
\begin{tikzpicture}[line cap=round,line join=round,scale=1.4]
  \draw (-5:0.8mm)\foreach\i in {2,...,25}{--(\i*30:\i*0.5mm)}--
    (775:13.5mm)\foreach\i in {25,...,1}{--(\i*30:\i*0.5mm+1mm)}--cycle;
  \draw[blue] (175:0.8mm)\foreach\i in {2,...,22}{--(\i*30+180:\i*0.5mm)}--
    (865:12mm)\foreach\i in {22,...,1}{--(\i*30+180:\i*0.5mm+1mm)}--cycle;
  \path (775:13.5mm)--(810:11.5mm) coordinate[pos=-0.7] (A) coordinate[pos=2.75] (B);
  \draw[dashed] (A)--(B);
  \path (865:12mm)--(540:10mm) coordinate[pos=-1.1] (C) coordinate[pos=2.6] (D);
  \draw[dashed] (C)--(D);
  \node[right] at (750:13.5mm) {$P_0$};
  \node[blue,above left] at (865:12mm) {$P_1$};
  \path (865:12mm)--(540:10mm) node[pos=0.5,left] {1};
  \path (775:13.5mm)--(810:11.5mm) node[pos=0.5,above] {2};
\end{tikzpicture}
\end{minipage}\\[2.5ex]
\begin{tabular}{r|c|c|c|c|}
& $\HHH_0(a,b)$ & $\HHH_1(a,b)$ & orientation of $P_0$ & orientation of $P_1$ \\
\hline
1 & $\RHP(a,b)$ & $\RHP(a,b)$ & counterclockwise & clockwise \\
\hline
2 & $\wideLHP(a,b)$ & $\wideLHP(a,b)$ & clockwise & counterclockwise \\
\hline
3 & $\RHP(a,b)$ & $\wideLHP(a,b)$ & clockwise & clockwise \\
\hline
4 & $\wideLHP(a,b)$ & $\RHP(a,b)$ & counterclockwise & counterclockwise \\
\hline
\end{tabular}
\caption{The meaning of $\HHH_0(a,b)$ and $\HHH_1(a,b)$ and the assumed orientations of $P_0$ and $P_1$ in the pseudocodes of Algorithm~\ref{ALG1} and Algorithm~\ref{ALG2}, depending on which common tangent of 1--4 is requested.}
\label{fig:assumptions}
\end{figure}

The particular case of 1--4 to be solved is specified to the algorithms by providing two binary parameters $\alpha_0,\alpha_1\in\{+1,-1\}$ specifying that the final solution $(s_0,s_1)$ should satisfy
\begin{align*}
P_0&\subset\RHP(p_0[s_0],p_1[s_1])\quad\text{if }\alpha_0=+1\text{,} & P_0&\subset\LHP(p_0[s_0],p_1[s_1])\quad\text{if }\alpha_0=-1\text{,} \\
P_1&\subset\RHP(p_0[s_0],p_1[s_1])\quad\text{if }\alpha_1=+1\text{,} & P_1&\subset\LHP(p_0[s_0],p_1[s_1])\quad\text{if }\alpha_1=-1\text{.}
\end{align*}
For clarity, instead of using the parameters $\alpha_0$ and $\alpha_1$ explicitly, the pseudocode uses half-planes $\HHH_0(a,b)$ and $\HHH_1(a,b)$ defined as follows, for any $k\in\{0,1\}$ and any distinct $a,b\in\setR^2$:
\begin{equation*}
\HHH_k(a,b)=\bigl\{c\in\setR^2\colon\alpha_k\dets(a,b,c)\leq 0\bigr\}=\begin{cases}
\RHP(a,b) & \text{if }\alpha_k=+1\text{,} \\
\wideLHP(a,b) & \text{if }\alpha_k=-1\text{.}
\end{cases}
\end{equation*}
The final solution $(s_0,s_1)$ should satisfy $P_k\subset\HHH_k(p_0[s_0],p_1[s_1])$.
A test of the form $c\notin\HHH_k(a,b)$ in the pseudocode should be understood as testing whether $\alpha_k\dets(a,b,c)>0$.
Another assumption that we make when presenting the pseudocode concerns the direction in which each polygon $P_k$ is traversed in order to find an index $v_k$ such that $p_k[v_k]\notin\HHH_k(p_0[s_0],p_1[s_1])$.
For a reason that will become clear later when we analyze correctness of the algorithms, we require that
\begin{itemize}
\item $P_0$ is traversed counterclockwise when $\alpha_1=+1$ and clockwise when $\alpha_1=-1$,
\item $P_1$ is traversed clockwise when $\alpha_0=+1$ and counterclockwise when $\alpha_0=-1$.
\end{itemize}
In the pseudocode, the forward orientation of $P_k$ is \emph{assumed} to be the one in which the corners of $P_k$ should be traversed according to the conditions above.
When this has not been guaranteed in the problem setup, a reference to a corner of $P_k$ of the form $p_k[i]$ in the pseudocode should be understood as $p_k[\beta_ki]$ for the constant $\beta_k\in\{+1,-1\}$ computed as follows at the very beginning:
\begin{align*}
\beta_0&=\alpha_1\sgn\dets(p_0[0],\ldots,p_0[n_0-1])\text{,} & \beta_1&=-\alpha_0\sgn\dets(p_1[0],\ldots,p_1[n_1-1])\text{.}
\end{align*}
The assumptions made in the pseudocode of the two algorithms for each particular case of 1--4 are summarized in Figure~\ref{fig:assumptions}.

Algorithm~\ref{ALG1} maintains a candidate solution $(s_0,s_1)$ starting from $(s_0,s_1)=(0,0)$.
At the beginning and after each update to $(s_0,s_1)$, the algorithm traverses $P_0$ and $P_1$ in parallel with indices $(v_0,v_1)$, starting from $(v_0,v_1)=(s_0,s_1)$ and advancing $v_0$ and $v_1$ alternately.
The variable $u\in\{0,1\}$ determines the polygon $P_u$ in which the traversal is advanced in the current iteration.
If the test in line~\ref{line1:side} of Algorithm~\ref{ALG1} succeeds, that is, the corner $p_u[v_u]$ lies on the ``wrong side'' of the candidate line, then the algorithm updates the candidate solution by setting $s_u\gets v_u$ and reverts $v_{1-u}$ back to $s_{1-u}$ in line~\ref{line1:update}.
The algorithm returns $(s_0,s_1)$ when both polygons have been entirely traversed with indices $v_0$ and $v_1$ without detecting any corner on the ``wrong side'' of the candidate line.
This can happen only when $P_0\subset\HHH_0(p_0[s_0],p_1[s_1])$ and $P_1\subset\HHH_1(p_0[s_0],p_1[s_1])$, as required.

\begin{figure}[t]
\centering
\begin{tikzpicture}[line cap=round,line join=round,scale=0.2]
  \draw[postaction={on each segment=mid arrow}] (21,22.5) coordinate (B)--(19.4,35.5)--(14.0,36.3)--(11.9,31.0)
    --(9.2,41.0)--(2.6,38.7)--(7.5,33.1)--(4.5,26.5)--(0.5,31)--(-2.5,24.5)
    --(5.5,19.5) coordinate (A)--(7,24.5)--(11.5,28)--(16,28.5)--cycle;
  \draw[blue,postaction={on each segment=mid arrow}] (26.7,20.4)--(36.5,18.5)--(42.9,25.2)--(38.0,28.1)--(44.7,31.2)
    --(46.4,37.1)--(42.9,41.3)--(33,38.5) coordinate (C)--(33.5,33.5)
    --(30.5,30)--(24,29) coordinate (D)--(28.0,26.4)--(33.4,26.2)--cycle;
  \draw[black!63,dashed] (C)--(A);
  \draw[black!63,dashed] (C)--(B);
  \path (B)--(D) coordinate[pos=-0.66] (X) coordinate[pos=2.85] (Y);
  \draw[dashed] (X)--(Y);
  \node at (7.2,37.4) {$P_0$};
  \node[blue] at (39.5,35) {$P_1$};
  \tikzstyle{every node}=[circle,draw,fill,minimum size=2pt,inner sep=0pt]
  \tikzstyle{every label}=[rectangle,draw=none,fill=none,label distance=3pt]
  \node[label=below:$a$] at (A) {};
  \node[label=below right:$b$] at (B) {};
  \node[label=below right:$c$] at (C) {};
  \node[label=above left:$d$] at (D) {};
\end{tikzpicture}
\caption{An example of how Algorithm~\ref{ALG1} finds the separating common tangent $\LLL(b,d)$ of $P_0$ and $P_1$ starting from $(p_0[0],p_1[0])=(a,c)$.
The segments $p_0[s_0]p_1[s_1]$ on intermediate candidate lines are also shown.}
\label{fig:run1}
\end{figure}
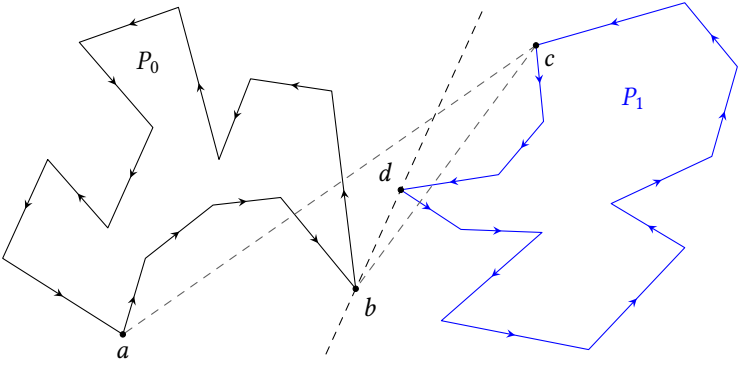

See Figure~\ref{fig:run1} for an example run of Algorithm~\ref{ALG1} for the separating common tangent problem (case~4).
The following theorem asserts that Algorithm~\ref{ALG1} is correct for the separating common tangent problem and ``partially correct'' for the outer common tangent problem.

\begin{theorem}
\label{thm:correctness1}
If Algorithm~\ref{ALG1} is to solve the outer common tangent problem (case 1 or~2), then it returns the solution\/ $(s_0,s_1)$ if the convex hulls of\/ $P_0$ and\/ $P_1$ are disjoint and reports ``no solution'' if the convex hulls of\/ $P_0$ and\/ $P_1$ are nested.
If Algorithm~\ref{ALG1} is to solve the separating common tangent problem (case 3 or~4), then it returns the solution\/ $(s_0,s_1)$ if the convex hulls of\/ $P_0$ and\/ $P_1$ are disjoint and reports ``no solution'' otherwise.
Moreover, Algorithm~\ref{ALG1} runs in linear time and uses constant workspace.
\end{theorem}

\begin{figure}[t]
\centering
\begin{tikzpicture}[line cap=round,line join=round,yscale=0.75]
  \coordinate (p0) at (5.4,0.1);
  \coordinate (p1) at (7.1,0.5);
  \coordinate (p2) at (6.5,1.6);
  \coordinate (p3) at (5.8,0.9);
  \coordinate (p4) at (4.3,1.7);
  \coordinate (p5) at (0.9,1.3);
  \coordinate (p6) at (1.3,4.6);
  \coordinate (p7) at (0.2,3.6);
  \coordinate (p8) at (1.1,6.3);
  \coordinate (p9) at (5,5.7);
  \coordinate (p10) at (4.5,7);
  \coordinate (p11) at (0.2,6.9);
  \coordinate (p12) at (-0.8,2.8);
  \coordinate (p13) at (0.4,0.4);
  \draw[postaction={on each segment=mid arrow}] (p0)--(p1)--(p2)--(p3)--(p4)--(p5)--(p6)--(p7)--(p8)--(p9)--(p10)--(p11)--(p12)--(p13)--cycle;
  \coordinate (q0) at (2.9,2.8);
  \coordinate (q1) at (4.5,3);
  \coordinate (q2) at (6.6,5.2);
  \coordinate (q3) at (7.3,2.3);
  \coordinate (q4) at (2.2,2.2);
  \coordinate (q5) at (2.4,5.1);
  \coordinate (q6) at (3.8,5.3);
  \coordinate (q7) at (4.9,4.6);
  \coordinate (q8) at (3.5,4.6);
  \coordinate (q9) at (4.1,3.9);
  \draw[blue,postaction={on each segment=mid arrow}] (q0)--(q1)--(q2)--(q3)--(q4)--(q5)--(q6)--(q7)--(q8)--(q9)--cycle;
  \draw[black!63,dashed] (p0)--(q0);
  \draw[black!63,dashed] (p0)--(q4);
  \draw[black!63,dashed] (p5)--(q4);
  \draw[black!63,dashed] (p5)--(q5);
  \draw[black!63,dashed] (p6)--(q5);
  \draw[black!63,dashed] (p8)--(q5);
  \draw[black!63,dashed] (p8)--(q6);
  \draw[black!63,dashed] (p9)--(q6);
  \draw[black!63,dashed] (p9)--(q7) coordinate[pos=-1.5] (a) coordinate[pos=5.4] (b);
  \draw[dotted] (a)--(p9) (q7)--(b);
  \draw[red!63,dashed] (p0)--(q7);
  \draw[red!63,dashed] (p0)--(q8);
  \draw[black!63,dashed] (p9)--(q2);
  \path (p10)--(q2) coordinate[pos=-0.3] (x) coordinate[pos=1.37] (y);
  \draw[dashed] (x)--(y);
  \node at (0.3,2.7) {$P_0$};
  \node[blue] at (6,3.2) {$P_1$};
  \tikzstyle{every node}=[circle,draw,fill,minimum size=2pt,inner sep=0pt]
  \tikzstyle{every label}=[rectangle,draw=none,fill=none,label distance=3pt]
  \node[label=below:$a$] at (p0) {};
  \node[label={[yshift=-2pt]above right:$b$}] at (p9) {};
  \node[label=above right:$c$] at (p10) {};
  \node[label=left:$d$] at (q0) {};
  \node[label=right:$e$] at (q7) {};
  \node[label={[yshift=-2pt]above right:$f$}] at (q2) {};
\end{tikzpicture}
\caption{An example of how Algorithm~\ref{ALG2} finds and Algorithm~\ref{ALG1} fails to find the outer common tangent $\LLL(c,f)$ of $P_0$ and $P_1$ starting from $(p_0[0],p_1[0])=(a,d)$.
The segments $p_0[s_0]p_1[s_1]$ on intermediate candidate lines are shown by dashed lines---gray for those considered by Algorithm~\ref{ALG2} and red for those considered by Algorithm~\ref{ALG1} but not Algorithm~\ref{ALG2}.
Both algorithms proceed along the same lines until the $26$th iteration.
In particular, in the $18$th iteration of both algorithms, an update makes $(p_0[s_0],p_1[s_1])=(b,e)$ and the dotted line $\LLL(b,e)$ becomes the candidate line.
In the $27$th iteration, after the assignment $v_u\gets v_u+1$, the algorithms encounter $u=0$ and $p_0[v_0]=a$.
Algorithm~\ref{ALG1} then makes $p_0[s_0]=a$, and in the $33$rd iteration, it reaches back the state $(p_0[s_0],p_1[s_1])=(p_0[v_0],p_1[v_1])=(a,d)$ and $u=0$ where it started except that the indices $s_0$, $v_0$, $s_1$, and $v_1$ have increased by the sizes of the respective polygons.
If the conditions $s_0<2n_0$ and $s_1<2n_1$ of the ``while'' loop were ignored, Algorithm~\ref{ALG1} would keep updating $(s_0,s_1)$ indefinitely.
By contrast, in the $27$th iteration of Algorithm~\ref{ALG2}, the test in line~\ref{line2:triangle} succeeds and $b_0$ is set.
Algorithm~\ref{ALG2} continues by making $(p_0[s_0],p_1[s_1])=(b,f)$ in the $28$th iteration (while clearing $b_0$) and finally $(p_0[s_0],p_1[s_1])=(c,f)$ in the $29$th iteration.}
\label{fig:run2}
\end{figure}
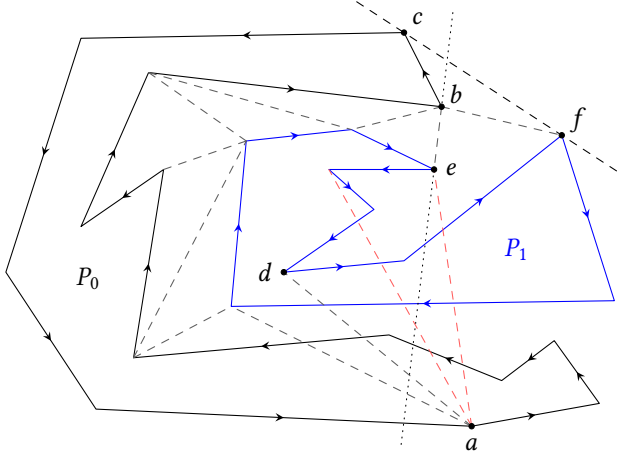

If the convex hulls of $P_0$ and $P_1$ properly overlap, then Algorithm~\ref{ALG1} can fail to find the solution even though it exists.
An example of such behavior is presented in Figure~\ref{fig:run2}.
Algorithm~\ref{ALG2} is an improved version of Algorithm~\ref{ALG1} that solves the problem correctly in all cases including the case of properly overlapping convex hulls.
In line~\ref{line2:triangle} of Algorithm~\ref{ALG2}, $\Delta(a,b,c)$ denotes the triangular region spanned by $a$, $b$, and $c$, and a test of the form $z\in\Delta(a,b,c)$ is equivalent to testing whether $\dets(z,a,b)$, $\dets(z,b,c)$, and $\dets(z,c,a)$ are all positive or all negative (they are all non-zero, by the general position assumption).
Suppose that $u=0$ (for simplicity of the explanation that follows) and the test in line~\ref{line2:triangle} succeeds.
Hence $p_1[s_1]$ belongs to the convex hull of $P_0$.
Algorithm~\ref{ALG1} would now set $s_0$ to $v_0$.
Intuitively, this would ``reverse'' the orientation of the candidate line (which is most evident when the angle $p_0[s_0]p_1[s_1]p_0[v_0]$ is close to $\pi$), possibly leading to failure.
Algorithm~\ref{ALG2} proceeds differently: $s_0$ remains unchanged, and a boolean variable $b_0$ is set in order to prevent updates to $s_0$ in further iterations of the algorithm until one of the iterations makes an update to $s_1$ and clears $b_0$ in line~\ref{line2:update}.
As we will show, such an update to $s_1$ must occur unless the convex hull of $P_1$ is contained in the convex hull of $P_0$, and preventing updates to $s_0$ when $b_0$ is set suffices to guarantee correctness of the algorithm in all cases.

See Figure~\ref{fig:run2} for an example run of Algorithm~\ref{ALG2} for the outer common tangent problem (case~1), where the convex hulls of $P_0$ and $P_1$ properly overlap.
If the convex hulls of $P_0$ and $P_1$ are disjoint, then the test in line~\ref{line2:triangle} of Algorithm~\ref{ALG2} never succeeds, the variables $b_0$ and $b_1$ remain unset, and thus Algorithm~\ref{ALG2} essentially becomes Algorithm~\ref{ALG1}.

\begin{theorem}
\label{thm:correctness2}
If Algorithm~\ref{ALG2} is to solve the outer common tangent problem (case 1 or~2), then it returns the solution\/ $(s_0,s_1)$ unless the convex hulls of\/ $P_0$ and\/ $P_1$ are nested, in which case it reports ``no solution''.
If Algorithm~\ref{ALG2} is to solve the separating common tangent problem (case 3 or~4), then it returns the solution\/ $(s_0,s_1)$ if the convex hulls of\/ $P_0$ and\/ $P_1$ are disjoint and reports ``no solution'' otherwise.
Moreover, Algorithm~\ref{ALG2} runs in linear time and uses constant workspace.
\end{theorem}

\section[Correctness of Algorithms \ref{ALG1} and~\ref{ALG2}]{Correctness of Algorithm~\ref{ALG1} and Algorithm~\ref{ALG2}}
\label{sec:correctness}

In this section, we prove Theorem~\ref{thm:correctness1} and Theorem~\ref{thm:correctness2} on correctness and efficiency of Algorithms \ref{ALG1} and~\ref{ALG2} while leaving the proof of a key lemma to the next section.
First, we prove the claims on running time and workspace usage in Theorems \ref{thm:correctness1} and~\ref{thm:correctness2}.

\begin{lemma}
\label{lem:timespace}
Algorithms \ref{ALG1} and~\ref{ALG2} run in linear time and use constant workspace.
\end{lemma}

\begin{proof}
It is clear that the algorithms use constant workspace.
For the bound on the running time, we prove the following two claims:
\begin{enumerate}
\item\label{item:difference} Before and after every iteration of the ``while'' loop in Algorithm \ref{ALG1} or~\ref{ALG2}, we have $(v_u-s_u)-(v_{1-u}-s_{1-u})\in\{-1,0\}$.
\item\label{item:sum} In each iteration, the sum $s_0+s_1+v_0+v_1$ is increased by at least~$1$.
\end{enumerate}
Initially, we have $s_u=v_u=s_{1-u}=v_{1-u}=0$, so statement~\ref{item:difference} holds before the first iteration.
Now, suppose that statement~\ref{item:difference} holds before iteration~$i$.
After the assignment $v_u\gets v_u+1$, we have $(v_u-s_u)-(v_{1-u}-s_{1-u})\in\{0,1\}$, and the sum $s_0+s_1+v_0+v_1$ has been increased by~$1$.
The former implies that if the assignments $s_u\gets v_u$ and $v_{1-u}\gets s_{1-u}$ are performed in iteration~$i$ (in line~\ref{line1:update} of Algorithm~\ref{ALG1} or line~\ref{line2:update} of Algorithm~\ref{ALG2}), then the sum $s_0+s_1+v_0+v_1$ remains unchanged or is increased by $1$ again, and we have $(v_u-s_u)-(v_{1-u}-s_{1-u})=0$ afterwards.
In total, the sum $s_0+s_1+v_0+v_1$ is increased by $1$ or $2$ in iteration~$i$.
Finally, after the assignment $u\gets 1-u$, we have $(v_u-s_u)-(v_{1-u}-s_{1-u})\in\{-1,0\}$, so statement~\ref{item:difference} holds at the end of iteration~$i$.

Statement~\ref{item:difference} implies that each iteration starts with $s_0<2n_0$, $s_1<2n_1$, $v_0-s_0\leq\max(n_0,n_1)$, and $v_1-s_1\leq\max(n_0,n_1)$ (where at least one of the last two inequalities is strict), otherwise the ``while'' loop would terminate before that iteration.
Therefore, we have $s_0+s_1+v_0+v_1\leq 2(s_0+s_1)+2\max(n_0,n_1)<6(n_0+n_1)$ before each iteration fully performed by the algorithm, in particular the last one.
This, the fact that $s_0+s_1+v_0+v_1=0$ before the first iteration, and statement~\ref{item:sum} imply that the algorithm makes at most $6(n_0+n_1)$ iterations, each of which takes constant time.
\end{proof}

Let $k\in\{0,1\}$.
We extend the notation $p_k[x]$ to all real numbers $x$ to make the function $\setR\ni x\mapsto p_k[x]\in P_k$ a continuous and piecewise linear traversal of $P_k$ wrapping around with period $n_k$:
\begin{equation*}
p_k[x]=(\lceil x\rceil-x)p_k[\lfloor x\rfloor]+(x-\lfloor x\rfloor)p_k[\lceil x\rceil]\in p_k[\lfloor x\rfloor]p_k[\lceil x\rceil]\text{,\quad for }x\in\setR\setminus\setZ\text{.}
\end{equation*}
When $x,y\in\setR$ and $x\leq y$, we let $P_k[x,y]$ denote the part of $P_k$ from $p_k[x]$ to $p_k[y]$ in the forward direction of $P_k$, that is, $P_k[x,y]=\{p_k[z]\colon z\in[x,y]\}$.
We say that $P_k[x,y]$ is a \emph{cap} of $\HHH_k(a,b)$ (for distinct points $a,b\in\setR^2$) if $P_k[x,y]\subset\HHH_k(a,b)$ and $P_k[x,y]\cap\LLL(a,b)=\{x,y\}$; this allows $x=y$.

\begin{lemma}
\label{lem:invariant1}
For each\/ $k\in\{0,1\}$, Algorithm~\ref{ALG1} maintains the following invariant before and after every iteration of the ``while'' loop:\/ $P_k[s_k,v_k]\subset\HHH_k(p_0[s_0],p_1[s_1])$.
\end{lemma}

\begin{proof}
Algorithm~\ref{ALG1} starts with $s_0=v_0=s_1=v_1=0$, so the invariant holds initially.
To show that it is preserved by every iteration, suppose it holds before iteration~$i$.
Let $u$ and $v_u$ denote the values in iteration~$i$ after the assignment $v_u\gets v_u+1$ and before the assignment $u\gets 1-u$.
Suppose the test in line~\ref{line1:side} of Algorithm~\ref{ALG1} succeeds, that is, $p_u[v_u]\notin\HHH_u(p_0[s_0],p_1[s_1])$; otherwise, clearly, the invariant is preserved by iteration~$i$.
The updates in line~\ref{line1:update} yield $s_0=v_0$ and $s_1=v_1$, which makes the invariant satisfied after iteration~$i$.
\end{proof}

\begin{figure}[t]
\centering
\begin{tikzpicture}[line cap=round,line join=round]
  \tikzstyle{every node}=[circle,draw,fill,minimum size=2pt,inner sep=0pt]
  \tikzstyle{every label}=[rectangle,draw=none,fill=none,label distance=3pt]
  \draw[dotted] (-3.4,0)--(3.9,0);
  \draw[red] (-0.9,0)--(-0.7,-0.9)--(2.7,-0.8)--(2.8,0.8) node[red,label=above:{$p_0[v_0]$}] {};
  \draw (-1.05,-0.8)--(-0.9,0)--(-1,0.8) node[label=above:{$p_0[v_0]$}] {};
  \draw (-0.9,0)--(-1.4,-0.7)--(-2.2,-0.6)--(-2.3,0.8) node[label=above:{$p_0[v_0]$}] {};
  \draw (-0.9,0)--(-0.5,-0.6)--(0.3,-0.5)--(0.3,0.8) node[label=above:{$p_0[v_0]$}] {};
  \node[label=above right:{$p_0[s_0]$}] at (-0.9,0) {};
  \draw[blue] (2,0) node[label=above:{$p_1[s_1]$}] {};
  \tikzstyle{every node}=[cross out,draw,fill=none,minimum size=3pt,inner sep=0pt]
  \tikzstyle{every label}=[rectangle,draw=none,fill=none,label distance=1pt]
  \draw (-2.243,0) node[label=below left:{$p_0[w_0]$}] {};
  \draw (0.3,0) node[label=below right:{$\,p_0[w_0]$}] {};
  \draw[red] (2.75,0) node[label=below right:{$p_0[w_0]$}] {};
  \path[on each segment=mid arrow] (-1.05,-0.8)--(-0.9,0)--(-1,0.8);
  \path[on each segment=mid arrow] (-1.4,-0.7)--(-2.2,-0.6);
  \path[on each segment=mid arrow] (-0.5,-0.6)--(0.3,-0.5);
  \path[on each segment={mid arrow,red}] (-0.7,-0.9)--(2.7,-0.8);
\end{tikzpicture}
\caption{Illustration for the statement of Lemma~\ref{lem:invariant2} in four possible cases of how $p_0[v_0-1]p_0[v_0]$ can intersect the candidate line $\LLL(p_0[s_0],p_1[s_1])$ when the test in line~\ref{line2:side} of Algorithm~\ref{ALG2} succeeds.
The half-plane $\HHH_0(p_0[s_0],p_1[s_1])$ is below the dotted line.
Algorithm~\ref{ALG2} makes the update $s_0\gets v_0$ in the first three cases.
In the last case (drawn red), $b_0$ becomes set and the second part of the invariant in Lemma~\ref{lem:invariant2} becomes satisfied.}
\label{fig:invariant2}
\end{figure}
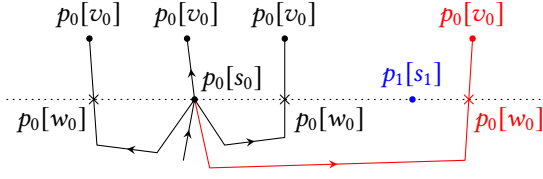

\begin{lemma}
\label{lem:invariant2}
{\upshape (See Figure~\ref{fig:invariant2} for an illustration.)}
For each\/ $k\in\{0,1\}$, Algorithm~\ref{ALG2} maintains the following invariant before and after every iteration of the ``while'' loop:
\begin{itemize}
\item if\/ $b_k=\FALSE$, then\/ $P_k[s_k,v_k]\subset\HHH_k(p_0[s_0],p_1[s_1])$;
\item if\/ $b_k=\TRUE$, then there is\/ $w_k\in(s_k,v_k)$ such that\/ $P_k[s_k,w_k]$ is a cap of\/ $\HHH_k(p_0[s_0],p_1[s_1])$ and\/ $p_{1-k}[s_{1-k}]\in p_k[s_k]p_k[w_k]$.
\end{itemize}
Moreover, on every update\/ $s_u\gets v_u$ in line~\ref{line2:update} of Algorithm~\ref{ALG2}, if\/ $s_u$ denotes the value before the update, then\/ $p_u[v_u]\notin\HHH_u(p_0[s_0],p_1[s_1])$ and there is\/ $w_u\in[v_u-1,v_u)$ such that\/ $P_u[s_u,w_u]$ is a cap of\/ $\HHH_u(p_0[s_0],p_1[s_1])$ and\/ $p_{1-u}[s_{1-u}]\notin p_u[s_u]p_u[w_u]$.
\end{lemma}

\begin{proof}
Algorithm~\ref{ALG2} starts with $s_0=v_0=s_1=v_1=0$ and $b_0=b_1=\FALSE$, so the invariant holds initially.
To show that it is preserved by every iteration, suppose it holds before iteration~$i$.
Let $u$ and $v_u$ denote the values in iteration~$i$ after the assignment $v_u\gets v_u+1$ and before the assignment $u\gets 1-u$.
Suppose the test in line~\ref{line2:side} of Algorithm~\ref{ALG2} succeeds, that is, $p_u[v_u]\notin\HHH_u(p_0[s_0],p_1[s_1])$ and $b_u=\FALSE$; otherwise, clearly, the invariant is preserved by iteration~$i$.
This and the assumption that the invariant holds before iteration~$i$ imply $P_u[s_u,v_u-1]\subset\HHH_u(p_0[s_0],p_1[s_1])$ and $P_u[s_u,v_u]\nsubset\HHH_u(p_0[s_0],p_1[s_1])$.
Let $w_u\in[v_u-1,v_u)$ be maximal such that $P_u[s_u,w_u]\subset\HHH_u(p_0[s_0],p_1[s_1])$.
That is, $p_u[w_u]$ is the intersection point of $p_u[v_u-1]p_u[v_u]$ and $\LLL(p_0[s_0],p_1[s_1])$.
The general position assumption implies that $P_u[s_u,w_u]$ is a cap of $\HHH_u(p_0[s_0],p_1[s_1])$.
We have $p_{1-u}[s_{1-u}]\in p_u[s_u]p_u[w_u]$ if and only if $p_{1-u}[s_{1-u}]\in\Delta(p_u[s_u],p_u[v_u-1],p_u[v_u])$.
Therefore, if the test in line~\ref{line2:triangle} succeeds, then the assignment $b_u\gets\TRUE$ in line~\ref{line2:setb} makes the invariant satisfied after iteration~$i$.
Now, suppose the test in line~\ref{line2:triangle} fails.
It follows that $p_{1-u}[s_{1-u}]\notin p_u[s_u]p_u[w_u]$, so the update $s_u\gets v_u$ in line~\ref{line2:update} satisfies the second statement of the lemma.
Furthermore, the updates in line~\ref{line2:update} yield $s_0=v_0$, $s_1=v_1$, and $b_{1-u}=\FALSE$, which makes the invariant satisfied after iteration~$i$.
\end{proof}

Most effort in proving correctness of the two algorithms lies in the following two lemmas:

\begin{lemma}
\label{lem:main1}
If the convex hulls of\/ $P_0$ and\/ $P_1$ are disjoint, then the ``while'' loop in Algorithm~\ref{ALG1} ends with\/ $s_0<2n_0$ and\/ $s_1<2n_1$.
\end{lemma}

\begin{lemma}
\label{MAINLEMMA}
If Algorithm~\ref{ALG2} is to solve the outer common tangent problem (case 1 or~2) and the convex hulls of\/ $P_0$ and\/ $P_1$ are not nested or Algorithm~\ref{ALG2} is to solve the separating common tangent problem (case 3 or~4) and the convex hulls of\/ $P_0$ and\/ $P_1$ are disjoint, then the ``while'' loop in Algorithm~\ref{ALG2} ends with\/ $s_0<2n_0$ and\/ $s_1<2n_1$.
\end{lemma}

If the convex hulls of $P_0$ and $P_1$ are disjoint, then the test in line~\ref{line2:triangle} of Algorithm~\ref{ALG2} never succeeds, $b_1$ and $b_2$ remain unset all the time, and thus Algorithm~\ref{ALG2} becomes equivalent to Algorithm~\ref{ALG1}.
Therefore, Lemma~\ref{lem:main1} is a direct consequence of Lemma~\ref{MAINLEMMA}.
We prove Lemma~\ref{MAINLEMMA} in the next section.
Here, we proceed with the proofs of Theorems \ref{thm:correctness1} and~\ref{thm:correctness2} assuming Lemma~\ref{MAINLEMMA}.

\begin{proof}[Proof of Theorem~\ref{thm:correctness1}]
Algorithm~\ref{ALG1} returns $(s_0,s_1)$ only when the ``while'' loop has terminated with $v_0\geq s_0+n_0$ and $v_1\geq s_1+n_1$, which implies $P_0\subset\HHH_0(p_0[s_0],p_1[s_1])$ and $P_1\subset\HHH_1(p_0[s_0],p_1[s_1])$, in view of Lemma~\ref{lem:invariant1}.
Therefore, whenever Algorithm~\ref{ALG1} returns $(s_0,s_1)$, it is the correct solution.
This implies that Algorithm~\ref{ALG1} correctly reports ``no solution'' in all cases where there is indeed no solution, that is, if the algorithm is to solve the outer common tangent problem (case 1 or~2) and the convex hulls of $P_0$ and $P_1$ are nested or it is to solve the separating common tangent problem (case 3 or~4) and the convex hulls of $P_0$ and $P_1$ are not disjoint.
By Lemma~\ref{lem:main1}, if the convex hulls of $P_0$ and $P_1$ are disjoint, then the ``while'' loop ends with $s_0<2n_0$ and $s_1<2n_1$, so the algorithm returns $(s_0,s_1)$, which is then the correct solution, as we have already argued.
By Lemma~\ref{lem:timespace}, the running time and the workspace usage are as stated.
\end{proof}

For the proof of correctness of Algorithm~\ref{ALG2}, we need one more lemma.

\begin{lemma}
\label{lem:b0b1false}
If Algorithm~\ref{ALG2} is to solve the outer common tangent problem (case 1 or~2) and the convex hulls of\/ $P_0$ and\/ $P_1$ are not nested or Algorithm~\ref{ALG2} is to solve the separating common tangent problem (case 3 or~4) and the convex hulls of\/ $P_0$ and\/ $P_1$ are disjoint, then the ``while'' loop in Algorithm~\ref{ALG2} ends with\/ $b_0=b_1=\FALSE$.
\end{lemma}

\begin{proof}
Consider the final values of $s_0$, $v_0$, $b_0$, $s_1$, $v_1$, and $b_1$ when the ``while'' loop in Algorithm~\ref{ALG2} is terminated.
Lemma~\ref{MAINLEMMA} yields $s_0<2n_0$ and $s_1<2n_1$.
This and the termination condition implies $v_0\geq s_0+n_0$ and $v_1\geq s_1+n_1$.
If the algorithm is to solve the separating common tangent problem (case 3 or~4) and we have $b_0=\TRUE$ or $b_1=\TRUE$, then Lemma~\ref{lem:invariant2} implies that the convex hulls of $P_0$ and $P_1$ are not disjoint, contrary to the assumption of the present lemma.
Now, suppose the algorithm is to solve the outer common tangent problem (case 1 or~2).
Thus $\HHH_0(p_0[s_0],p_1[s_1])=\HHH_1(p_0[s_0],p_1[s_1])$.
If $b_k=\TRUE$ and $b_{1-k}=\FALSE$ for some $k\in\{0,1\}$, then Lemma~\ref{lem:invariant2} and the fact that $P_{1-k}[s_{1-k},v_{1-k}]=P_{1-k}$ yield a cap $P_k[s_k,w_k]$ of $\HHH_k(p_0[s_0],p_1[s_1])$ such that $P_{1-k}$ is contained in the polygonal region bounded by $P_k[s_k,w_k]\cup p_k[s_k]p_k[w_k]$, and therefore the convex hull of $P_{1-k}$ is contained in the convex hull of $P_k$, contrary to the assumption of the lemma.
If $b_0=b_1=\TRUE$, then Lemma~\ref{lem:invariant2} yields a cap $P_0[s_0,w_0]$ of $\HHH_0(p_0[s_0],p_1[s_1])$ and a cap $P_1[s_1,w_1]$ of $\HHH_1(p_0[s_0],p_1[s_1])$ such that the points $p_1[w_1]$, $p_0[s_0]$, $p_1[s_1]$, and $p_0[w_0]$ occur in this order on $\LLL(p_0[s_0],p_1[s_1])$, which is impossible when $\HHH_0(p_0[s_0],p_1[s_1])=\HHH_1(p_0[s_0],p_1[s_1])$.
\end{proof}

\begin{proof}[Proof of Theorem~\ref{thm:correctness2}]
Algorithm~\ref{ALG2} returns $(s_0,s_1)$ only when the ``while'' loop has terminated with $v_0\geq s_0+n_0$, $v_1\geq s_1+n_1$, and $b_0=b_1=\FALSE$,
which implies $P_0\subset\HHH_0(p_0[s_0],p_1[s_1])$ and $P_1\subset\HHH_1(p_0[s_0],p_1[s_1])$, in view of Lemma~\ref{lem:invariant2}.
Therefore, whenever Algorithm~\ref{ALG2} returns $(s_0,s_1)$, it is the correct solution.
This implies that Algorithm~\ref{ALG2} correctly reports ``no solution'' in all cases where there is indeed no solution, that is, if the algorithm is to solve the outer common tangent problem (case 1 or~2) and the convex hulls of $P_0$ and $P_1$ are nested or it is to solve the separating common tangent problem (case 3 or~4) and the convex hulls of $P_0$ and $P_1$ are not disjoint.
In the other cases, the ``while'' loop ends with $s_0<2n_0$ and $s_1<2n_1$, by Lemma~\ref{MAINLEMMA}, and with $b_0=b_1=\FALSE$, by Lemma~\ref{lem:b0b1false}, and therefore the algorithm returns $(s_0,s_1)$, which is then the correct solution, as we have already argued.
By Lemma~\ref{lem:timespace}, the running time and the workspace usage are as stated.
\end{proof}

\section{Proof of Lemma~\ref{MAINLEMMA}}
\label{sec:main-lemma}

To complete the proof of correctness of the two algorithms, it remains to prove Lemma~\ref{MAINLEMMA}.
For the rest of this section, we adopt the assumptions of Lemma~\ref{MAINLEMMA}, in particular the assumption that the solution exists, and we show that it is found and returned by Algorithm~\ref{ALG2}.

\subsection[Reduction to one case]{Reduction to one case of the outer common tangent problem}

We will reduce Lemma~\ref{MAINLEMMA} for all cases 1--4 of the common tangent problem just to case~1.
First, we explain what we mean by such a reduction.
An input to Algorithm~\ref{ALG2} is a quadruple $(P_0,P_1,\alpha_0,\alpha_1)$, where $P_0=\PPP(p_0[0],\ldots,p_0[n_0-1])$, $P_1=\PPP(p_1[0],\ldots,p_1[n_1-1])$, and $\alpha_0,\alpha_1\in\{+1,-1\}$ are implicit parameters that determine the particular case 1--4 of the common tangent problem to be solved by the algorithm (see Section~\ref{sec:description}).
Consider two inputs $(P_0,P_1,\alpha_0,\alpha_1)$ and $(P'_0,P'_1,\alpha'_0,\alpha'_1)$, where $P_k=\PPP(p_k[0],\ldots,p_k[n_k-1])$ and $P'_k=\PPP(p'_k[0],\ldots,p'_k[n_k-1])$ for each $k\in\{0,1\}$.
Whenever $a$ (or $b$, $c$, etc.) denotes the point $p_k[i]$ (where $k\in\{0,1\}$ and $i\in\setZ$), we let $a'$ (or $b'$, $c'$, etc.) denote the point $p'_k[i]$.
The inputs $(P_0,P_1,\alpha_0,\alpha_1)$ and $(P'_0,P'_1,\alpha'_0,\alpha'_1)$ are \emph{equivalent} if the following holds for all $a,b,c\in\{p_0[0],\ldots,p_0[n_0-1],p_1[0],\ldots,p_1[n_1-1]\}$:
\begin{align*}
\alpha_0\sgn\dets(a,b,c)&=\alpha'_0\sgn\dets(a',b',c')\quad\text{if }\lvert\{a,b,c\}\cap\{p_0[0],\ldots,p_0[n_0-1]\}\rvert\in\{0,2\}\text{,} \\
\alpha_1\sgn\dets(a,b,c)&=\alpha'_1\sgn\dets(a',b',c')\quad\text{if }\lvert\{a,b,c\}\cap\{p_0[0],\ldots,p_0[n_0-1]\}\rvert\in\{1,3\}\text{.}
\end{align*}
With this definition, equivalent inputs have the same solutions and lead to the same outcomes of Algorithm~\ref{ALG2}, as we show in the next two lemmas.
Therefore, equivalence of inputs provides a formal way of reducing one case of Lemma~\ref{MAINLEMMA} to another one.

\begin{lemma}
\label{lem:equiv-solution}
If inputs\/ $(P_0,P_1,\alpha_0,\alpha_1)$ and\/ $(P'_0,P'_1,\alpha'_0,\alpha'_1)$ are equivalent and\/ $s_0,s_1\in\setZ$, then\/ $(s_0,s_1)$ either is the correct solution or is not the correct solution to both inputs.
\end{lemma}

\begin{proof}
The conditions in the definition of equivalence imply that $\alpha_k\dets(p_0[s_0],p_1[s_1],p_k[i])>0$ if and only if $\alpha'_k\dets(p'_0[s_0],p'_1[s_1],p'_k[i])>0$, for any $k\in\{0,1\}$ and $i\in\setZ$.
This implies that $P_k\subset\HHH_k(p_0[s_0],p_1[s_1])$ if and only if $P'_k\subset\HHH_k(p'_0[s_0],p'_1[s_1])$, for any $k\in\{0,1\}$, that is, $\LLL(p_0[s_0],p_1[s_1])$ is the requested common tangent of $P_0$ and $P_1$ if and only if $\LLL(p'_0[s_0],p'_1[s_1])$ is the requested common tangent of $P'_0$ and $P'_1$.
\end{proof}

\begin{lemma}
\label{lem:equiv-algorithm}
If inputs\/ $(P_0,P_1,\alpha_0,\alpha_1)$ and\/ $(P'_0,P'_1,\alpha'_0,\alpha'_1)$ are equivalent, then Algorithm~\ref{ALG2} applied to\/ $(P_0,P_1,\alpha_0,\alpha_1)$ and\/ $(P'_0,P'_1,\alpha'_0,\alpha'_1)$ ends with the same final value of the pair of variables\/ $(s_0,s_1)$.
\end{lemma}

\begin{proof}
Let $k\in\{0,1\}$.
Recall from Section~\ref{sec:description} the implicit constant $\beta_k\in\{+1,-1\}$, which determines whether Algorithm~\ref{ALG2} traverses $P_k$ ($P'_k$) forwards or backwards.
We show that $\beta_k$ has the same value for both inputs.
It is well known that $P_k$ can be \emph{triangulated}; in particular, there are $n_k-2$ triangles of the form $\PPP(a_t,b_t,c_t)$ with $a_t,b_t,c_t\in\{p_k[0],\ldots,p_k[n_k-1]\}$ ($1\leq t\leq n_k-2$), all with the same orientation (counterclockwise or clockwise), such that
\begin{equation*}
\dets(p_k[0],\ldots,p_k[n_k-1])=\sum_{t=1}^{n-2}\bigl(\det(a_t,b_t)+\det(b_t,c_t)+\det(c_t,a_t)\bigr)=\sum_{t=1}^{n-2}\dets(a_t,b_t,c_t)\text{.}
\end{equation*}
Equivalence of $(P_0,P_1,\alpha_0,\alpha_1)$ and $(P'_0,P'_1,\alpha'_0,\alpha'_1)$ implies
\begin{equation*}
\alpha_{1-k}\sgn\dets(a_t,b_t,c_t)=\alpha'_{1-k}\sgn\dets(a'_t,b'_t,c'_t)\quad\text{for all }t\in\{1,\ldots,n_k-2\}\text{.}
\end{equation*}
We conclude that all triangles $\PPP(a'_t,b'_t,c'_t)$ ($1\leq t\leq n_k-2$) have the same orientation and
\begin{equation*}
\alpha_{1-k}\sgn\dets(p_k[0],\ldots,p_k[n_k-1])=\alpha'_{1-k}\sgn\dets(p'_k[0],\ldots,p'_k[n_k-1])\text{.}
\end{equation*}
This and the definition of $\beta_k$ implies that $\beta_k$ has the same value for both inputs.

We show that Algorithm~\ref{ALG2} proceeds in exactly the same way for both inputs.
Specifically, we show the same number of iterations of the ``while'' loop is performed for both inputs, and for each iteration~$i$, the variables $s_0$, $v_0$, $b_0$, $s_1$, $v_1$, and $b_1$ have the same values for both inputs before and after iteration~$i$ (it is clear that $u$ has the same value, because it depends only on $i$).
This implies, in particular, that the algorithm ends with the same final value of the pair of variables $(s_0,s_1)$.

The initial setup is common for both inputs.
Now, suppose $s_0$, $v_0$, $b_0$, $s_1$, $v_1$, and $b_1$ have the same values for both inputs before iteration~$i$.
The test in line \ref{line2:side} produces the same outcome for both inputs, because equivalence of $(P_0,P_1,\alpha_0,\alpha_1)$ and $(P'_0,P'_1,\alpha'_0,\alpha'_1)$ implies that $\alpha_u\dets(p_0[s_0],p_1[s_1],p_u[v_u])>0$ if and only if $\alpha'_u\dets(p'_0[s_0],p'_1[s_1],p'_u[v_u])>0$.
If that outcome is positive, then the test in line \ref{line2:triangle} produces the same outcome for both inputs, by an analogous argument.
It follows that the same assignments are performed in iteration~$i$ for both inputs, and therefore $s_0$, $v_0$, $b_0$, $s_1$, $v_1$, and $b_1$ have the same values for both inputs after iteration~$i$.
\end{proof}

First, we reduce Lemma~\ref{MAINLEMMA} for cases 2 and~4 of the common tangent problem to cases 1 and~3 thereof.
Consider the transformation $\phi\colon\setR^2\ni(x,y)\mapsto(-x,y)\in\setR^2$ (horizontal flip).
Let $P'_0=\PPP(\phi(p_0[0]),\ldots,\phi(p_0[n_0-1]))$ and $P'_1=\PPP(\phi(p_1[0]),\ldots,\phi(p_1[n_1-1]))$.
Clearly, for any three points $a,b,c\in\setR^2$, we have $\dets(\phi(a),\phi(b),\phi(c))=-\dets(a,b,c)$.
It follows that the input $(P_0,P_1,\alpha_0,\alpha_1)$ is equivalent to $(P'_0,P'_1,-\alpha_0,-\alpha_1)$.
If the former is case 2 or~4 of the common tangent problem, then the latter is case 1 or~3 thereof, respectively.
By Lemmas \ref{lem:equiv-solution} and~\ref{lem:equiv-algorithm}, it remains to prove Lemma~\ref{MAINLEMMA} for cases 1 and~3 of the common tangent problem.

Now, we reduce Lemma~\ref{MAINLEMMA} for case~3 of the common tangent problem to case~1 thereof.
Suppose an input $(P_0,P_1,\alpha_0,\alpha_1)$ is case~3 of the common tangent problem, that is, $\alpha_0=+1$ and $\alpha_1=-1$.
Assume that the convex hulls of $P_0$ and $P_1$ are disjoint (as in Lemma~\ref{MAINLEMMA}).
It follows that there is a straight line separating the two convex hulls in the plane.
Assume without loss of generality that it is the vertical line $x=0$ and every corner of $P_0$ has negative $x$-coordinate while every corner of $P_1$ has positive $x$-coordinate, applying an appropriate rotation or translation of the plane to turn $(P_0,P_1,\alpha_0,\alpha_1)$ into an equivalent input that has these properties.
Consider the transformation
\begin{equation*}
\phi\colon(\setR\setminus\{0\})\times\setR\ni(x,y)\mapsto\left(\frac{y}{x},\frac{1}{x}\right)\in(\setR\setminus\{0\})\times\setR\text{.}
\end{equation*}
For any three points $a=(a_x,a_y)$, $b=(b_x,b_y)$, and $c=(c_x,c_y)$ in $(\setR\setminus\{0\})\times\setR$, we have
\begin{equation}
\tag{$\dagger$}\label{eq:dets}
\dets(\phi(a),\phi(b),\phi(c))=\begin{vmatrix}
  \frac{a_y}{a_x} & \frac{b_y}{b_x} & \frac{c_y}{c_x} \\[2pt]
  \frac{1}{a_x} & \frac{1}{b_x} & \frac{1}{c_x} \\
  1 & 1 & 1
\end{vmatrix}=\begin{vmatrix}
  1 & 1 & 1 \\
  \frac{a_y}{a_x} & \frac{b_y}{b_x} & \frac{c_y}{c_x} \\[2pt]
  \frac{1}{a_x} & \frac{1}{b_x} & \frac{1}{c_x}
\end{vmatrix}=\frac{\dets(a,b,c)}{a_xb_xc_x}\text{.}
\end{equation}
Since $a$, $b$, and $c$ are collinear if and only if $\dets(a,b,c)=0$, it follows from \eqref{eq:dets} that $\phi$ preserves collinearity (actually, it is a projective transformation).
This and the fact that $\phi$ is a bijection on $(\setR\setminus\{0\})\times\setR$ imply that $\phi$ transforms the polygons $P_0$ and $P_1$ into (simple) polygons $P'_0=\PPP(\phi(p_0[0]),\ldots,\phi(p_0[n_0-1]))$ and $P'_1=\PPP(\phi(p_1[0]),\ldots,\phi(p_1[n_1-1]))$, respectively.
Since the corners of $P_0$ have negative $x$-coordinates and the corners of $P_1$ have positive $x$-coordinates, the equality \eqref{eq:dets} implies the following, for all $a,b,c\in\{p_0[0],\ldots,p_0[n_0-1],p_1[0],\ldots,p_1[n_1-1]\}$:
\begin{alignat*}{2}
\sgn\dets(\phi(a),\phi(b),\phi(c))&={}&\sgn\dets(a,b,c)\quad&\text{if }\lvert\{a,b,c\}\cap\{p_0[0],\ldots,p_0[n_0-1]\}\rvert\in\{0,2\}\text{,} \\
\sgn\dets(\phi(a),\phi(b),\phi(c))&={}&-\sgn\dets(a,b,c)\quad&\text{if }\lvert\{a,b,c\}\cap\{p_0[0],\ldots,p_0[n_0-1]\}\rvert\in\{1,3\}\text{.}
\end{alignat*}
Therefore, the inputs $(P_0,P_1,\alpha_0,\alpha_1)$ and $(P'_0,P'_1,\alpha_0,-\alpha_1)$ are equivalent.
Since the former is case~3 of the common tangent problem, the latter is case~1 thereof.
By Lemmas \ref{lem:equiv-solution} and~\ref{lem:equiv-algorithm}, it remains to prove Lemma~\ref{MAINLEMMA} for case~1 of the problem.
This is what the remainder of Section~\ref{sec:main-lemma} is devoted to.

\subsection{Auxiliary concepts}
\label{sec:auxiliary}

For the sequel, we assume that the convex hulls of $P_0$ and $P_1$ are not nested, $P_0$ is oriented counterclockwise, $P_1$ is oriented clockwise, and Algorithm~\ref{ALG2} is to solve case~1 of the outer common tangent problem---compute a pair of indices $(s_0,s_1)$ such that $P_0,P_1\subset\RHP(p_0[s_0],p_1[s_1])$.

Recall that a segment $ab$ in the plane is considered oriented from $a$ to $b$, so that forward traversal of $ab$ starts at $a$ and ends at $b$.
A \emph{polygonal path} is a curve in the plane composed of $n$ segments $a_0a_1,a_1a_2,\ldots,a_{n-1}a_n$ with no common points other than the common endpoints of pairs of consecutive segments in that order.
Such a polygonal path is considered oriented from $a_0$ to $a_n$, so that forward traversal of it starts at $a_0$ and ends at $a_n$.
A segment or a polygonal path is \emph{degenerate} if it consists of a single point.
When a non-degenerate polygonal path $a_0a_1,a_1a_2,\ldots,a_{n-1}a_n$ (a non-degenerate segment if $n=1$) is contained in the boundary of a polygonal region $Q$, we say that $Q$ lies \emph{to the left} or \emph{to the right} of $a_0a_1,a_1a_2,\ldots,a_{n-1}a_n$ if forward traversal of $a_0a_1,a_1a_2,\ldots,a_{n-1}a_n$ agrees with counterclockwise traversal or clockwise traversal, respectively, of the boundary of $Q$.
For $k\in\{0,1\}$ and $a,b\in P_k$, let $P_k[a,b]$ be the polygonal path from $a$ to $b$ along $P_k$ in the forward direction (counterclockwise for $P_0$ and clockwise for $P_1$); in particular, $P_k[a,a]=\{a\}$.

Let $\ell_0,r_0\in P_0$ and $\ell_1,r_1\in P_1$ be such that $P_0,P_1\subset\LHP(\ell_0,\ell_1)\cap\RHP(r_0,r_1)$.
Thus $r_0$ and $r_1$ determine the requested outer common tangent, while $\ell_0$ and $\ell_1$ determine the other one.
It is possible that $\ell_0=r_0$ or $\ell_1=r_1$ (but not both).
For clarity of presentation, we will ignore this special case and proceed as if $\ell_0\neq r_0$ and $\ell_1\neq r_1$.
Our arguments remain correct when $\ell_k=r_k$ ($k\in\{0,1\}$) after adding the following exceptions to the definitions of a polygon, a polygonal path, and $P_k[a,b]$:
\begin{itemize}
\item the point $\ell_k=r_k$ is allowed to occur twice on a polygon (as two corners) or a polygonal path (as both endpoints of the path), where one occurrence is denoted by $\ell_k$ and the other by $r_k$;
\item $P_k[\ell_k,r_k]=P_k$; forward traversal of $P_k[\ell_k,r_k]$ makes one full traversal of $P_k$ from $\ell_k$ to $r_k$ in the forward direction of $P_k$ (counterclockwise for $P_0$ and clockwise for $P_1$).
\end{itemize}

\begin{figure}[t]
\centering
\begin{tikzpicture}[line cap=round,line join=round,scale=0.5,yscale=1.1]
  \path (6.46,5.26)--(2.26,3.52) coordinate[pos=0.3] (A);
  \path (8.54,5.32)--(10.56,6.96) coordinate[pos=0.7] (B);
  \path (12.12,2.96)--(9.66,5.1) coordinate[pos=0.7] (C);
  \path (9.66,5.1)--(8.48,2.32) coordinate[pos=0.75] (D);
  \path (5.34,6.26)--(5.26,9.04) coordinate[pos=0.3] (E);
  \path (6.46,3.14)--(2.26,3.52) coordinate[pos=0.25] (F);
  \fill[black!10] (1.16,1.24)--(4.44,2.08)--(2.52,2.44)--(6.46,3.14)--(2.26,3.52)--(6.46,5.26)--(5.34,6.26)--(5.26,9.04)
    --(10.16,9.26)--(8,7.6)--(8.54,5.32)--(10.56,6.96)--(12.5,6.7)--(12.12,2.96)--(9.66,5.1)--(8.48,2.32)--(11,1.24)--cycle;
  \fill[red!25] (6.46,3.14)--(A)--(6.46,5.26)--(8,7.6)--(8.54,5.32)--(B)--(C)--(9.66,5.1)--(D)--cycle;
  \draw (2.54,8.12)--(2.84,6.22)--(4.45,5.5)--(4.02,6.6)--(5.26,9.04)--(5.34,6.26)
    --(6.46,5.26)--(2.26,3.52)--(6.46,3.14)--(2.52,2.44)--(4.44,2.08)--(1.16,1.24)
    --(0.2,4.02)--(3.68,4.74)--(1.68,5.04)--(0.52,7.74)--(1.7,7.48)--cycle;
  \draw[blue] (10.16,9.26)--(8,7.6)--(8.54,5.32)--(10.56,6.96)--(12.5,6.7)--(12.12,2.96)
    --(9.66,5.1)--(8.48,2.32)--(11,1.24)--(14.54,2.14)--(16.04,4.9)--(13.5,2.94)
    --(13.6,7.12)--(16,6.88)--(14.8,8.5)--(11.22,7.72)--(12.64,8.7)--cycle;
  \draw[dashed] (1.16,1.24) node[below] {$\ell_0$}--(11,1.24) node[below] {$\ell_1$};
  \draw[dashed] (5.26,9.04) node[above] {$r_0$}--(10.16,9.26) node[above] {$r_1$};
  \draw[dashed] (6.46,3.14) node[below,yshift=-1pt] {$x$}--(D) node[right,yshift=-2pt] {$y$};
  \draw[dashed] (8,7.6) node[above left,xshift=4pt,yshift=-1pt] {$y'$}--(6.46,5.26) node[right,xshift=-2pt,yshift=-4pt] {$x'$};
  \draw[dashed] (8,7.6)--(E);
  \draw[dashed] (8.54,5.32)--(F);
  \draw[dotted] (6.46,3.14)--(A);
  \draw[dotted] (B)--(C);
  \node[black!60] at (6.2,8.2) {\Large $E$};
  \node[red] at (7.9,4) {$Z$};
  \path (1.68,5.04)--(0.52,7.74) node[pos=0.2,left] {$P_0$};
  \path (13.5,2.94)--(13.6,7.12) node[pos=0.6,blue,right] {$P_1$};
\end{tikzpicture}
\caption{Illustration for the concepts of a door and a zone and for Observations \ref{obs:between}--\ref{obs:zone}.
Some pairwise non-crossing doors including $\ell_0\ell_1$ and $r_0r_1$ are indicated by dashed lines.
The polygonal regions $P_0[\ell_0,r_0]$ and $P_1[\ell_1,r_1]$ and the doors $\ell_0\ell_1$ and $r_0r_1$ determine the shaded zone $E$, which contains all doors and zones.
The boundary of the red zone $Z$ contains two doors $xy\prec x'y'$; $Z$ lies to the left of $xy$ and to the right of $x'y'$.}
\label{fig:auxiliary}
\end{figure}
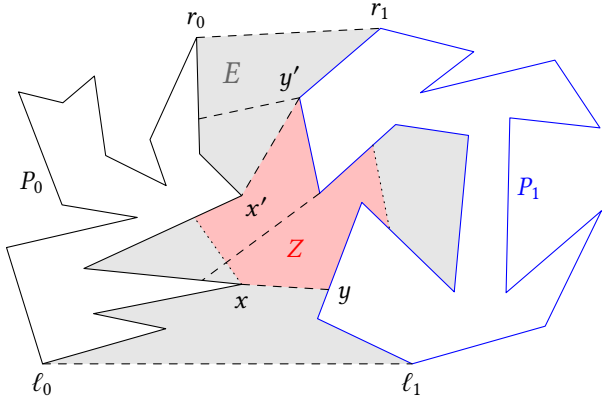

A \emph{door} is a segment $xy$ such that $xy\cap P_0=\{x\}$ and $xy\cap P_1=\{y\}$.
We always orient the door from the endpoint on $P_0$ to the endpoint on $P_1$.
A \emph{zone} is a polygonal region $Z$ such that the interior of $Z$ is disjoint from $P_0\cup P_1$ and the boundary of $Z$ is the union of some non-empty part of $P_0$ (not necessarily connected), some non-empty part of $P_1$ (likewise), and some segments with both endpoints on $P_0\cup P_1$ (not necessarily doors).
These concepts are illustrated in Figure~\ref{fig:auxiliary}.
Let $E$ be the polygonal region bounded by $\ell_0\ell_1\cup P_0[\ell_0,r_0]\cup P_1[\ell_1,r_1]\cup r_0r_1$.
Since $E\subset\LHP(\ell_0,\ell_1)\cap\RHP(r_0,r_1)$, it follows that $E$ lies to the left of $\ell_0\ell_1$, to the right of $P_0[\ell_0,r_0]$, to the left of $P_1[\ell_1,r_1]$, and to the right of $r_0r_1$.
Figure~\ref{fig:auxiliary} also illustrates the next three observations.

\begin{observation}
\label{obs:between}
The polygonal region\/ $E$ is a zone and satisfies\/ $E\cap P_0=P_0[\ell_0,r_0]$ and\/ $E\cap P_1=P_1[\ell_1,r_1]$.
Moreover, every door or zone is contained in\/ $E$.
In particular, every door has one endpoint on\/ $P_0[\ell_0,r_0]$ and the other on\/ $P_1[\ell_1,r_1]$.
\end{observation}

\begin{proof}
Let $k\in\{0,1\}$ and $S_k=(\LHP(\ell_0,\ell_1)\cap\RHP(r_0,r_1))\setminus P_k[\ell_k,r_k]$.
Since $E$ and the polygonal region bounded by $P_k$ lie on opposite sides of $P_k[\ell_k,r_k]$, the sets $P_k\setminus P_k[\ell_k,r_k]$ and $E\setminus P_k[\ell_k,r_k]$ are contained in different connected components of $S_k$.
A consequence of this property is that $E\cap P_k=P_k[\ell_k,r_k]$.
This, for both $k\in\{0,1\}$, proves the first statement.
Another consequence is that $P_k\setminus P_k[\ell_k,r_k]$ and $P_{1-k}$ belong to different connected components of $S_k$, because $E\setminus P_k[\ell_k,r_k]$ and $P_{1-k}$ intersect.
Therefore, $P_k[\ell_k,r_k]$ intersects the interior of every segment or polygonal region that is contained in $\LHP(\ell_0,\ell_1)\cap\RHP(r_0,r_1)$ and intersects both $P_k\setminus P_k[\ell_k,r_k]$ and $P_{1-k}$.
However, every door or zone is contained in $\LHP(\ell_0,\ell_1)\cap\RHP(r_0,r_1)$ (because $P_0$ and $P_1$ are) and is internally disjoint from $P_k[\ell_k,r_k]$.
This, for both $k\in\{0,1\}$, proves the last two statements.
\end{proof}

For two doors $xy$ and $x'y'$, let $xy\preceq x'y'$ denote that $P_0[x,x']\subseteq P_0[\ell_0,r_0]$ and $P_1[y,y']\subseteq P_1[\ell_1,r_1]$ (that is, forward traversal of $P_0[\ell_0,r_0]$ encounters $x$ no later than $x'$ and forward traversal of $P_1[\ell_1,r_1]$ encounters $y$ no later than $y'$), and let $xy\prec x'y'$ denote that $xy\preceq x'y'$ and $xy\neq x'y'$.
Two doors are \emph{non-crossing} if they are disjoint or they intersect only at a common endpoint.
The following observation implies that $\prec$ is a total order on any set of pairwise non-crossing doors:

\begin{observation}
\label{obs:order}
Any two non-crossing doors\/ $xy$ and\/ $x'y'$ satisfy\/ $xy\prec x'y'$ or\/ $x'y'\prec xy$.
\end{observation}

\begin{proof}
Observation~\ref{obs:between} yields $xy,x'y'\subset E$.
If neither $xy\prec x'y'$ nor $x'y'\prec xy$, then the points $x$, $x'$, $y$, and $y'$ are distinct and occur on the boundary of $E$ in this cyclic order (clockwise or counterclockwise), which contradicts the assumption that $xy$ and $x'y'$ do not cross.
\end{proof}

\begin{observation}
\label{obs:zone}
The boundary of every zone\/ $Z$ contains exactly two doors.
Moreover, if these doors are denoted by\/ $xy$ and\/ $x'y'$ so that\/ $xy\prec x'y'$, then
\begin{enumerate}
\item\label{item:zone-side} $Z$ lies to the left of\/ $xy$ and to the right of\/ $x'y'$,
\item\label{item:zone-door} a door\/ $x''y''$ is disjoint from the interior of\/ $Z$ if and only if\/ $x''y''\preceq xy$ or\/ $x'y'\preceq x''y''$.
\end{enumerate}
\end{observation}

\begin{proof}
The boundary of $Z$ intersects both $P_0$ and $P_1$, so it contains at least two doors.
By Observation~\ref{obs:order}, since the doors on the boundary of $Z$ are pairwise non-crossing, they are totally ordered by $\prec$.
Let $xy$ be the minimum door and $x'y'$ be the maximum door on the boundary of $Z$ with respect to the order $\prec$.
We will show statements \ref{item:zone-side} and \ref{item:zone-door} for $xy$ and $x'y'$.
Statement \ref{item:zone-door}, minimality of $xy$, and maximality of $x'y'$ imply that the boundary of $Z$ contains no other doors.

For every door $x''y''$, since $Z\subseteq E$ and $x''y''\subset E$ (by Observation~\ref{obs:between}), the following holds: if the set $E\setminus x''y''$ has two connected components intersecting $Z$, then $x''y''$ intersects the interior of $Z$.
If neither $x''y''\preceq xy$ nor $x'y'\preceq x''y''$, then the set $E\setminus x''y''$ has two connected components intersecting $Z$ (by the definition of $\preceq$), so $x''y''$ intersects the interior of $Z$, which is one implication in statement~\ref{item:zone-door}.
It also follows that either of the sets $E\setminus xy$ and $E\setminus x'y'$ has only one connected component intersecting $Z$, so $Z$ is contained in the polygonal region $Z^\star$ bounded by $xy\cup P_0[x,x']\cup P_1[y,y']\cup x'y'$, which is contained in $E$.
If $x\neq x'$, then $Z^\star$ lies to the right of $P_0[x,x']$ (because $E$ does), and if $y\neq y'$, then $Z^\star$ lies to the left of $P_1[y,y']$ (because $E$ does).
Therefore, $Z^\star$ and thus $Z$ lie to the left of $xy$ and to the right of $x'y'$, which is statement~\ref{item:zone-side}.
Moreover, if $x''y''\preceq xy$ or $x'y'\preceq x''y''$, then $x''y''$ is disjoint from the interior of $Z^\star$ and therefore is disjoint from the interior of $Z$, which is the converse implication in statement~\ref{item:zone-door}.
\end{proof}

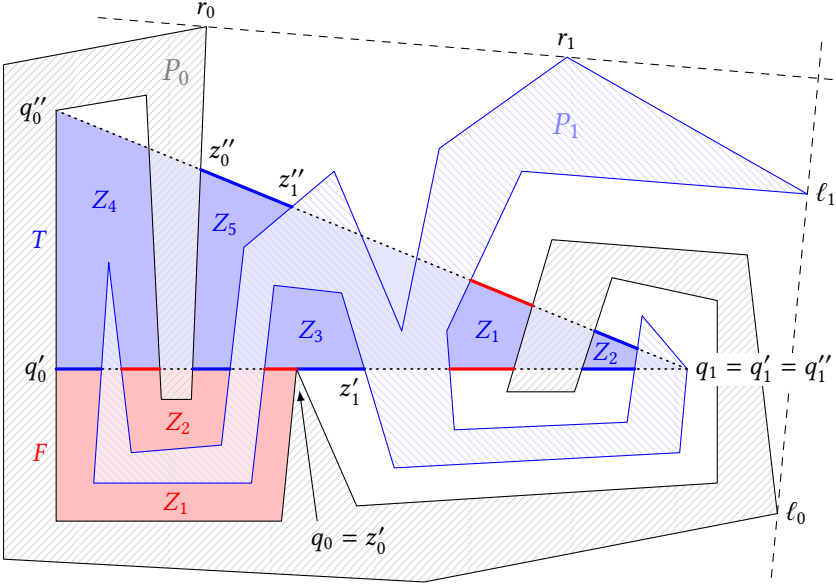
\begin{figure}[t]
\centering
\begin{tikzpicture}[line cap=round,line join=round]
  \tikzstyle{every node}=[rectangle,inner sep=3pt]
  \coordinate (P0) at (0,0) {};
  \coordinate (P1) at (-0.2,-2);
  \coordinate (P2) at (-3.2,-2);
  \coordinate (P3) at (-3.2,3.4);
  \coordinate (P4) at (-2,3.6);
  \coordinate (P5) at (-1.8,-0.4);
  \coordinate (P6) at (-1.4,-0.4);
  \coordinate (P7) at (-1.2,4.5);
  \coordinate (P8) at (-3.9,4);
  \coordinate (P9) at (-3.9,-2.5);
  \coordinate (P10) at (1.7,-2.8);
  \coordinate (P11) at (6.4,-1.9);
  \coordinate (P12) at (6,1.5);
  \coordinate (P13) at (3.4,1.7);
  \coordinate (P14) at (2.8,-0.3);
  \coordinate (P15) at (3.7,-0.3);
  \coordinate (P16) at (4.2,1.2);
  \coordinate (P17) at (5.6,0.9);
  \coordinate (P18) at (5.6,-1.5);
  \coordinate (P19) at (0.8,-1.8);
  \coordinate (Q0) at (5.2,0);
  \coordinate (Q1) at (5.1,-1.1);
  \coordinate (Q2) at (1.3,-1.3);
  \coordinate (Q3) at (0.6,1) {};
  \coordinate (Q4) at (-0.3,1.1) {};
  \coordinate (Q5) at (-0.6,-1.5);
  \coordinate (Q6) at (-2.7,-1.5);
  \coordinate (Q7) at (-2.5,1.4);
  \coordinate (Q8) at (-2.2,-1.1);
  \coordinate (Q9) at (-1,-1);
  \coordinate (Q10) at (-0.7,1.6);
  \coordinate (Q11) at (0.5,2.6);
  \coordinate (Q12) at (1.4,0.5);
  \coordinate (Q13) at (1.9,2.9);
  \coordinate (Q14) at (3.6,4.1);
  \coordinate (Q15) at (6.8,2.3);
  \coordinate (Q16) at (3,2.6);
  \coordinate (Q17) at (2,0.5);
  \coordinate (Q18) at (2.1,-0.8);
  \coordinate (Q19) at (4.4,-0.7);
  \coordinate (Q20) at (4.6,0.7);
  \coordinate (A5) at (-3.2,0);
  \coordinate (A1) at (intersection of Q0--A5 and P13--P14);
  \coordinate (B1) at (intersection of Q0--A5 and Q17--Q18);
  \coordinate (A2) at (intersection of Q0--A5 and P15--P16);
  \coordinate (B2) at (intersection of Q0--A5 and Q19--Q20);
  \coordinate (B3) at (intersection of Q0--A5 and Q2--Q3);
  \coordinate (B4) at (intersection of Q0--A5 and Q4--Q5);
  \coordinate (B5) at (intersection of Q0--A5 and Q6--Q7);
  \coordinate (A6) at (intersection of Q0--A5 and P4--P5);
  \coordinate (B6) at (intersection of Q0--A5 and Q7--Q8);
  \coordinate (A7) at (intersection of Q0--A5 and P6--P7);
  \coordinate (B7) at (intersection of Q0--A5 and Q9--Q10);
  \coordinate (C1) at (intersection of Q0--P3 and P13--P14);
  \coordinate (D1) at (intersection of Q0--P3 and Q16--Q17);
  \coordinate (C2) at (intersection of Q0--P3 and P15--P16);
  \coordinate (D2) at (intersection of Q0--P3 and Q19--Q20);
  \coordinate (C3) at (intersection of Q0--P3 and P6--P7);
  \coordinate (D3) at (intersection of Q0--P3 and Q10--Q11);
  \coordinate (C4) at (intersection of Q0--P3 and P4--P5);
  \path (P11)--(Q15) coordinate[pos=-0.2] (U0) coordinate[pos=1.48] (V0);
  \path (P7)--(Q14) coordinate[pos=-0.3] (U1) coordinate[pos=1.75] (V1);
  \draw[dashed] (U0)--(V0);
  \draw[dashed] (U1)--(V1);
  \fill[red!10] (P0)--(P1)--(P2)--(A5)--cycle;
  \fill[blue!10] (A5)--(P3)--(Q0)--cycle;
  \fill[red!25] (P0)--(P1)--(P2)--(A5)--(B5)--(Q6)--(Q5)--(B4)--cycle;
  \fill[red!25] (A6)--(P5)--(P6)--(A7)--(B7)--(Q9)--(Q8)--(B6)--cycle;
  \fill[blue!25] (A1)--(C1)--(D1)--(Q17)--(B1)--cycle;
  \fill[blue!25] (A2)--(C2)--(D2)--(B2)--cycle;
  \fill[blue!25] (B3)--(Q3)--(Q4)--(B4)--cycle;
  \fill[blue!25] (A5)--(P3)--(C4)--(A6)--(B6)--(Q7)--(B5)--cycle;
  \fill[blue!25] (A7)--(C3)--(D3)--(Q10)--(B7)--cycle;
  \draw[pattern=north east lines,pattern color=black!15] (P0)\foreach\i in {1,...,19}{--(P\i)}--cycle;
  \draw[blue,pattern=north west lines,pattern color=blue!15] (Q0)\foreach\i in {1,...,20}{--(Q\i)}--cycle;
  \draw[semithick,dotted] (Q0)--(A5);
  \draw[semithick,dotted] (Q0)--(P3);
  \draw[red,very thick] (A1)--(B1);
  \draw[blue,very thick] (A2)--(B2);
  \draw[blue,very thick] (P0)--(B3);
  \draw[red,very thick] (P0)--(B4);
  \draw[blue,very thick] (A5)--(B5);
  \draw[red,very thick] (A6)--(B6);
  \draw[blue,very thick] (A7)--(B7);
  \draw[red,very thick] (C1)--(D1);
  \draw[blue,very thick] (C2)--(D2);
  \draw[blue,very thick] (C3)--(D3);
  \node[inner sep=2pt] (L) at (0.7,-2.25) {$q_0=z'_0$};
  \draw[-latex,shorten >=8pt] (L.149)--(P0);
  \node[left] at (A5) {$q'_0$};
  \node[left] at (P3) {$q''_0$};
  \node[right,fill=white,inner sep=1pt,xshift=2pt] at (Q0) {$q_1=q'_1=q''_1$};
  \node[below left,xshift=2pt] at (B3) {$z'_1$};
  \node[above right,yshift=-2pt] at (C3) {$z''_0$};
  \node[above,yshift=1pt] at (D3) {$z''_1$};
  \node[right] at (P11) {$\ell_0$};
  \node[right] at (Q15) {$\ell_1$};
  \node[above] at (P7) {$r_0$};
  \node[above] at (Q14) {$r_1$};
  \node[left,red] at (-3.2,-1.1) {$F$};
  \node[left,blue] at (-3.2,1.7) {$T$};
  \node[red] at (-1.6,-1.77) {$Z_1$};
  \node[red] at (-1.57,-0.72) {$Z_2$};
  \node[blue] at (2.55,0.5) {$Z_1$};
  \node[blue] at (4.12,0.2) {$Z_2$};
  \node[blue] at (0.2,0.5) {$Z_3$};
  \node[blue] at (-2.55,2.2) {$Z_4$};
  \node[blue] at (-0.95,1.9) {$Z_5$};
  \node[black!50] at (-1.6,3.9) {\Large $P_0$};
  \node[blue!50] at (3.6,3.2) {\Large $P_1$};
\end{tikzpicture}
\caption{Thick blue segments are positive doors and thick red segments are negative doors on $q'_0q'_1$ and $q''_0q''_1$.
The primary doors are $z'_0z'_1$ on $q'_0q'_1$ and $z''_0z''_1$ on $q''_0q''_1$.
The red region $F$ (considered in Lemma~\ref{lem:jump}) determines red zones $Z_1\prec Z_2$.
The blue region $T$ (considered in Lemma~\ref{lem:move}) determines blue zones $Z_1\prec Z_2\prec Z_3\prec Z_4\prec Z_5$, of which $Z_1$, $Z_2$, and $Z_5$ are two-sided while $Z_3$ and $Z_4$ are one-sided.}
\label{fig:doors}
\end{figure}

For the rest of this subsection, fix points $q_0\in P_0$ and $q_1\in P_1$, and consider doors contained in $q_0q_1$ (\emph{doors on\/ $q_0q_1$} in short).
Recall that a door $xy$ is always oriented from its endpoint $x$ on $P_0$ to its endpoint $y$ on $P_1$.
The \emph{sign} of such a door $xy$ on $q_0q_1$ is
\begin{itemize}
\item $+1$ if forward traversal of $q_0q_1$ encounters $x$ first and $y$ second (then $xy$ is \emph{positive} on $q_0q_1$),
\item $-1$ if forward traversal of $q_0q_1$ encounters $y$ first and $x$ second (then $xy$ is \emph{negative} on $q_0q_1$).
\end{itemize}
See Figure~\ref{fig:doors} for an illustration.

\begin{observation}
\label{obs:signs}
The signs of all doors on\/ $q_0q_1$ sum up to\/~$1$.
\end{observation}

\begin{proof}
Let $x_1y_1,\ldots,x_dy_d$ be all the doors on $q_0q_1$ enumerated in the order they are encountered by forward traversal of $q_0q_1$.
The \emph{first endpoint} of such a door is the one closer to $q_0$, and the \emph{last endpoint} is the one closer to $q_1$.
Every subsegment of $q_0q_1$ connecting a point on $P_0$ with a point on $P_1$ contains at least one of the doors.
Since $q_0\in P_0$, the subsegment of $q_0q_1$ from $q_0$ to the first endpoint of $x_1y_1$ contains no points of $P_1$, so $x_1y_1$ is positive on $q_0q_1$.
For $i\in\{1,\ldots,d-1\}$, the subsegment of $q_0q_1$ from the last endpoint of $x_iy_i$ to the first endpoint of $x_{i+1}y_{i+1}$ contains no points of $P_0$ or no points of $P_1$, so the sign of $x_{i+1}y_{i+1}$ is opposite to the sign of $x_iy_i$ on $q_0q_1$.
Finally, since $q_1\in P_1$, the subsegment of $q_0q_1$ from the last endpoint of $x_dy_d$ to $q_1$ contains no points of $P_0$, so $x_dy_d$ is positive on $q_0q_1$.
This implies that $x_iy_i$ is positive on $q_0q_1$ for $i$ odd, $x_iy_i$ is negative on $q_0q_1$ for $i$ is even, and $d$ is odd.
Therefore, the signs of $x_1y_1,\ldots,x_dy_d$ on $q_0q_1$ sum up to~$1$.
\end{proof}

The doors on $q_0q_1$ are pairwise non-crossing, so they are totally ordered by the relation $\prec$, by Observation~\ref{obs:order}.
Let $x_1y_1,\ldots,x_dy_d$ be all the doors on $q_0q_1$ ordered so that $x_1y_1\prec\cdots\prec x_dy_d$.
The \emph{primary door} on $q_0q_1$ is the door $x_jy_j$ with minimum $j\in\{1,\ldots,d\}$ such that the signs of $x_1y_1,\ldots,x_jy_j$ on $q_0q_1$ sum up to~$1$.
Such an index $j$ exists, because $d$ is a candidate, by Observation~\ref{obs:signs}.
Minimality of $j$ in the definition of the primary door directly implies the following:

\begin{observation}
\label{obs:primary}
The primary door\/ $x_iy_i$ is positive on\/ $q_0q_1$.
Moreover, if\/ $i\geq 2$, then the door\/ $x_{i-1}y_{i-1}$ is also positive on\/ $q_0q_1$.
\end{observation}

Tracking the primary door on the segment $q_0q_1=p_0[s_0]p_1[s_1]$ as it changes in the course of the algorithm is the key idea in the proof of the remaining case of Lemma~\ref{MAINLEMMA} that follows.

\subsection[Proof for the remaining case]{Proof of Lemma~\ref{MAINLEMMA} for the remaining case}

We go back to the proof of Lemma~\ref{MAINLEMMA}.
Having reduced Lemma~\ref{MAINLEMMA} to case~1 of the outer common tangent problem, we have assumed the setup of that case: the convex hulls of $P_0$ and $P_1$ are not nested, $P_0$ is oriented counterclockwise, $P_1$ is oriented clockwise, and Algorithm~\ref{ALG2} is to compute a pair of indices $(s_0,s_1)$ such that $P_0,P_1\subset\RHP(p_0[s_0],p_1[s_1])$, that is, $p_0[s_0]=r_0$ and $p_1[s_1]=r_1$.

Algorithm~\ref{ALG2} starts with $(s_0,s_1)=(0,0)$ and then makes some updates to the candidate solution $(s_0,s_1)$ in line~\ref{line2:update} until the end of the ``while'' loop.
The second part of Lemma~\ref{lem:invariant2} explains what these updates look like: on every update $s_u\gets v_u$ in line~\ref{line2:update} of Algorithm~\ref{ALG2}, if $s_u$ denotes the value before the update, then $p_u[v_u]\notin\RHP(p_0[s_0],p_1[s_1])$ and there is $w_u\in[v_u-1,v_u)$ such that $P_u[s_u,w_u]$ is a cap of $\RHP(p_0[s_0],p_1[s_1])$, $P_u[w_u,v_u]$ is the segment $p_u[w_u]p_u[v_u]$, and $p_{1-u}[s_{1-u}]\notin p_u[s_u]p_u[w_u]$.

First, we present informally the general proof idea.
Imagine that an update like above happens in continuous time, as follows.
Let $q_0=p_0[s_0]$ and $q_1=p_1[s_1]$.
If $s_u=w_u$, then the point $q_u$ moves continuously along the segment $p_u[w_u]p_u[v_u]$ from $p_u[w_u]$ to $p_u[v_u]$.
If $s_u<w_u$, then the point $q_u$ jumps over all points $p_u[x]$ with $x\in(s_u,w_u]$ and then moves continuously along the segment $p_u[w_u]p_u[v_u]\setminus\{p_u[w_u]\}$ as in the case $s_u=w_u$.
Thus $q_u=p_u[s_u]$ again after the assignment $s_u\gets v_u$.
As $q_u$ is moving during the update, we track the primary door $z_0z_1$ on $q_0q_1$ and show that
\begin{enumerate}
\item\label{item:door-forward} the door $z_0z_1$ is only moving (piecewise continuously) forward in the order $\prec$,
\item\label{item:no-jump-over} the point $q_u$ never passes or jumps over $z_u$.
\end{enumerate}
To see how this implies Lemma~\ref{MAINLEMMA}, consider the overall move of $q_0$, $q_1$, and $z_0z_1$ during all updates to the candidate solution $(s_0,s_1)$, starting from $q_0=p_0[0]$ and $q_1=p_1[0]$.
For each $k\in\{0,1\}$, statement~\ref{item:door-forward} implies that $z_k$ is only moving forward on $P_k[\ell_k,r_k]$, never passing or jumping over $r_k$, and statement~\ref{item:no-jump-over} asserts that $q_k$ never passes or jumps over $z_k$, whence it follows that $q_k$ passes or jumps over $r_k$ at most once.

Now, we proceed with the proof of Lemma~\ref{MAINLEMMA}.
The next two lemmas formalize statement~\ref{item:door-forward} above---Lemma~\ref{lem:jump} for the initial jump over $P_u[s_u,w_u]$, and Lemma~\ref{lem:move} for the continuous move along $p[w_u]p[v_u]$.
See Figure~\ref{fig:doors} for an illustration of Lemmas \ref{lem:jump} and~\ref{lem:move}.
Statement~\ref{item:no-jump-over} above is formalized by the invariant in Lemma~\ref{lem:no-jump-over}.

\begin{lemma}
\label{lem:jump}
Let\/ $u\in\{0,1\}$, $q_u,q'_u\in P_u$, and\/ $q_{1-u}=q'_{1-u}\in P_{1-u}$.
If\/ $P_u[q_u,q'_u]$ is a cap of\/ $\RHP(q_0,q_1)$ and\/ $q_{1-u}\notin q_uq'_u$, then the same door is primary on\/ $q_0q_1$ and on\/ $q'_0q'_1$.
\end{lemma}

\begin{proof}
The lemma is trivial when $q_u=q'_u$, so assume $q_u\neq q'_u$.
Thus $q_0q_1\subset q'_0q'_1$ or $q'_0q'_1\subset q_0q_1$ (because $q_{1-u}\notin q_uq'_u$).
Let $q''_0q''_1$ be the longer of $q_0q_1$ and $q'_0q'_1$ ($q'_0q'_1$ in the former and $q_0q_1$ in the latter case).
Let $F$ be the polygonal region bounded by $P_u[q_u,q'_u]\cup q_uq_u'$.
Thus $F\subset\RHP(q''_0,q''_1)$.
Let $\ZZZ$ be the set of zones contained in $F$ with boundaries contained in $(P_0\cap F)\cup(P_1\cap F)\cup q_uq'_u$.
For every door $xy\subset q_uq'_u$, there is a zone in $\ZZZ$ to the right of $xy$ (if $F$ is to the right of $xy$) or to the left of $xy$ (if $F$ is to the left of $xy$).
By Observation~\ref{obs:zone}, the zones in $\ZZZ$ can be ordered as $Z_1,\ldots,Z_d$ and the doors contained in $q_uq'_u$ can be ordered as $x_1y_1,x^1y^1,\ldots,x_dy_d,x^{\smash[t]{d}}y^{\smash[t]{d}}$ so that
\begin{itemize}
\item every zone $Z_i\in\ZZZ$ has exactly two doors on the boundary, namely, $x_iy_i$ and $x^iy^i$,
\item $x_1y_1\prec x^1y^1\prec\cdots\prec x_dy_d\prec x^{\smash[t]{d}}y^{\smash[t]{d}}$.
\end{itemize}
For each $i\in\{1,\ldots,d\}$, since $Z_i\subset\RHP(q''_0,q''_1)$, Observation \ref{obs:zone} (\ref{item:zone-side} and~\ref{item:zone-door}) implies that
\begin{itemize}
\item $x_iy_i$ is negative and $x^iy^i$ is positive on $q''_0q''_1$,
\item $x_iy_i$ and $x^iy^i$ are consecutive in the order $\prec$ of the doors on $q''_0q''_1$.
\end{itemize}
By Observation~\ref{obs:primary}, none of $x_1y_1,x^1y^1,\ldots,x_dy_d,x^{\smash[t]{d}}y^{\smash[t]{d}}$ is primary on $q''_0q''_1$.
Moreover, for each door $xy$ on the shorter of $q_0q_1$ and $q'_0q'_1$, the following two sums are equal:
\begin{itemize}
\item the sum of the signs of all doors on $q''_0q''_1$ up to $xy$ in the order $\prec$,
\item the sum of the signs of all doors on the shorter of $q_0q_1$ and $q'_0q'_1$ up to $xy$ in the order $\prec$.
\end{itemize}
We conclude that the same door is primary on $q''_0q''_1$ and on the shorter of $q_0q_1$ and $q'_0q'_1$.
\end{proof}

\begin{lemma}
\label{lem:move}
Let\/ $u\in\{0,1\}$, $q'_uq''_u\subset P_u$, and\/ $q'_{1-u}=q''_{1-u}\in P_{1-u}$.
Let\/ $z'_0z'_1$ be the primary door on\/ $q'_0q'_1$ and\/ $z''_0z''_1$ be the primary door on\/ $q''_0q''_1$.
If\/ $q''_u\notin\RHP(q'_0,q'_1)$, then\/ $z'_0z'_1\prec z''_0z''_1$.
\end{lemma}

\begin{proof}
Let $T$ be the triangular region bounded by $q'_0q'_1\cup q'_0q''_0\cup q'_1q''_1\cup q''_0q''_1$, where either $q'_0q''_0$ or $q'_1q''_1$ is a degenerate segment.
Thus $T\subset\LHP(q'_0,q'_1)\cap\RHP(q''_0,q''_1)$.
Let $\ZZZ$ be the set of zones contained in $T$ with boundaries contained in $q'_0q'_1\cup(P_0\cap T)\cup(P_1\cap T)\cup q''_0q''_1$.
For every door $xy\subset q'_0q'_1\cup q''_0q''_1$, there is a zone in $\ZZZ$ to the right of $xy$ (if $T$ lies to the right of $xy$) or to the left of $xy$ (if $T$ lies to the left of $xy$).
By Observation~\ref{obs:zone}, the zones in $\ZZZ$ can be ordered as $Z_1,\ldots,Z_d$ and the doors contained in $q'_0q'_1\cup q''_0q''_1$ can be ordered as $x_1y_1,x^1y^1,\ldots,x_dy_d,x^{\smash[t]{d}}y^{\smash[t]{d}}$ so that
\begin{itemize}
\item every zone $Z_i\in\ZZZ$ has exactly two doors on the boundary, namely, $x_iy_i$ and $x^iy^i$,
\item $x_1y_1\prec x^1y^1\prec\cdots\prec x_dy_d\prec x^{\smash[t]{d}}y^{\smash[t]{d}}$.
\end{itemize}
For every $i\in\{1,\ldots,d\}$, since $Z_i\subset\LHP(q'_0,q'_1)\cap\RHP(q''_0,q''_1)$, Observation \ref{obs:zone}~(\ref{item:zone-side}) implies that
\begin{itemize}
\item $x_iy_i$ is a positive door on $q'_0q'_1$ or a negative door on $q''_0q''_1$,
\item $x^iy^i$ is a negative door on $q'_0q'_1$ or a positive door on $q''_0q''_1$.
\end{itemize}
We will use these two properties extensively without explicit reference.

We say that a zone $Z_i\in\ZZZ$ is \emph{one-sided} if $x_iy_i$ and $x^iy^i$ lie both on $q'_0q'_1$ or both on $q''_0q''_1$, otherwise we say that $Z_i$ is \emph{two-sided}.
For each one-sided zone $Z_i\in\ZZZ$, the doors $x_iy_i$ and $x^iy^i$ have opposite signs on $q'_0q'_1$ or $q''_0q''_1$ (whichever they lie on).
For each two-sided zone $Z_i\in\ZZZ$, if $x'_{\smash[t]{i}}y'_{\smash[t]{i}}$ and $x''_{\smash[t]{i}}y''_{\smash[t]{i}}$ denote the two doors on the boundary of $Z_i$ so that $x'_{\smash[t]{i}}y'_{\smash[t]{i}}\subseteq q'_0q'_1$ and $x''_{\smash[t]{i}}y''_{\smash[t]{i}}\subseteq q''_0q''_1$, then the sign of $x'_{\smash[t]{i}}y'_{\smash[t]{i}}$ on $q'_0q'_1$ is equal to the sign of $x''_{\smash[t]{i}}y''_{\smash[t]{i}}$ on $q''_0q''_1$.
Let $I$ be the set of indices $i\in\{1,\ldots,d\}$ such that $Z_i$ is a two-sided zone in $\ZZZ$.
The above and Observation~\ref{obs:signs} implies that
\begin{itemize}
\item the signs of the doors $x'_{\smash[t]{i}}y'_{\smash[t]{i}}$ on $q'_0q'_1$ over all $i\in I$ sum up to~$1$,
\item the signs of the doors $x''_{\smash[t]{i}}y''_{\smash[t]{i}}$ on $q''_0q''_1$ over all $i\in I$ sum up to~$1$,
\end{itemize}
and the following four sums are equal, for each $j\in I$:
\begin{itemize}
\item the sum of the signs of all doors on $q'_0q'_1$ up to $x'_{\smash[t]{j}}y'_{\smash[t]{j}}$ in the order $\prec$,
\item the sum of the signs of the doors $x'_{\smash[t]{i}}y'_{\smash[t]{i}}$ on $q'_0q'_1$ over all $i\in I\cap\{1,\ldots,j\}$,
\item the sum of the signs of the doors $x''_{\smash[t]{i}}y''_{\smash[t]{i}}$ on $q''_0q''_1$ over all $i\in I\cap\{1,\ldots,j\}$,
\item the sum of the signs of all doors on $q''_0q''_1$ up to $x''_{\smash[t]{j}}y''_{\smash[t]{j}}$ in the order $\prec$.
\end{itemize}
Let $j\in I$ be minimum such that the four sums above are equal to~$1$.
Since $x_iy_i$ is positive and $x^iy^i$ is negative on $q'_0q'_1$ for every (one-sided) zone $Z_i\in\ZZZ$ such that $x_iy_i,x^iy^i\subseteq q'_0q'_1$, it follows that the primary door on $q'_0q'_1$ is either $x'_{\smash[t]{j}}y'_{\smash[t]{j}}$ or $x_iy_i$ for some (one-sided) zone $Z_i\in\ZZZ$ with $x_iy_i,x^iy^i\subseteq q'_0q'_1$ and $i<j$, and thus $x_iy_i\prec x'_{\smash[t]{j}}y'_{\smash[t]{j}}$.
For every (one-sided) zone $Z_i\in\ZZZ$ such that $x_iy_i,x^iy^i\subseteq q''_0q''_1$, since $x_iy_i$ is negative and $x^iy^i$ is positive on $q''_0q''_1$, neither $x_iy_i$ nor $x^iy^i$ is primary on $q''_0q''_1$, by Observation~\ref{obs:primary}.
It follows that $x''_{\smash[t]{j}}y''_{\smash[t]{j}}$ is the primary door on $q''_0q''_1$.
Since the primary door is always positive, we have $x'_{\smash[t]{j}}y'_{\smash[t]{j}}=x_jy_j$ and $x''_{\smash[t]{j}}y''_{\smash[t]{j}}=x^jy^j$, and thus $x'_{\smash[t]{j}}y'_{\smash[t]{j}}\prec x''_{\smash[t]{j}}y''_{\smash[t]{j}}$.
We conclude that $z'_0z'_1\preceq x'_{\smash[t]{j}}y'_{\smash[t]{j}}\prec x''_{\smash[t]{j}}y''_{\smash[t]{j}}=z''_0z''_1$.
\end{proof}

\begin{lemma}
\label{lem:no-jump-over}
Let\/ $a_0\in[0,n_0)$ and\/ $a_1\in[0,n_1)$ be such that\/ $p_0[a_0]=\ell_0$ and\/ $p_1[a_1]=\ell_1$.
Algorithm~\ref{ALG2} maintains the following invariant: if\/ $c_0\in[a_0,a_0+n_0)$ and\/ $c_1\in[a_1,a_1+n_1)$ are such that\/ $p_0[c_0]p_1[c_1]$ is the primary door on\/ $p_0[s_0]p_1[s_1]$, then\/ $s_0\leq c_0$ and\/ $s_1\leq c_1$.
\end{lemma}

The invariant in Lemma~\ref{lem:no-jump-over} implies that $s_0\leq c_0<a_0+n_0<2n_0$ and $s_1\leq c_1<a_1+n_1<2n_1$ at the end of the ``while'' loop in Algorithm~\ref{ALG2}, which is the assertion of Lemma~\ref{MAINLEMMA}.

\begin{proof}
Algorithm~\ref{ALG2} starts with $s_0=0\leq a_0\leq c_0$ and $s_1=0\leq a_1\leq c_1$, so the invariant holds initially.
Consider an update $s_u\gets v_u$ in line~\ref{line2:update} of Algorithm~\ref{ALG2}, where $u\in\{0,1\}$, assuming that the invariant holds before the update.
Let $s_u$ denote the value before the update and $w_u$ be as claimed by the second part of Lemma~\ref{lem:invariant2}.
Let $q_u=p_u[s_u]$, $q'_u=p_u[w_u]$, $q''_u=p_u[v_u]$, and $q_{1-u}=q'_{1-u}=q''_{1-u}=p_{1-u}[s_{1-u}]$.
The conclusions of Lemma~\ref{lem:invariant2} imply that $q_uq'_u$ is a segment of $\LLL(q_0,q_1)$ not containing $q_{1-u}$, $P_u[q_u,q'_u]$ is a cap of $\RHP(q_0,q_1)$, $P_u[q'_u,q''_u]$ is the single segment $q'_uq''_u$, and $q''_u\notin\RHP(q_0,q_1)=\RHP(q'_0,q'_1)$, where the last equality follows from $q'_{1-u}=q_{1-u}\in\LLL(q_0,q_1)\setminus q_uq'_u$.
These conditions are what we need to apply Lemma~\ref{lem:jump} (to $q_0$, $q_1$, $q'_0$, and $q'_1$) and Lemma~\ref{lem:move} (to $q'_0$, $q'_1$, $q''_0$, and $q''_1$).
Let $z_0z_1$, $z'_0z'_1$, and $z''_0z''_1$ be the primary doors on $q_0q_1$, $q'_0q'_1$, and $q''_0q''_1$, respectively.
Lemma~\ref{lem:jump} and Lemma~\ref{lem:move} yield $z_0z_1=z'_0z'_1\prec z''_0z''_1$.
Let $c_0,c''_0\in[a_0,a_0+n_0)$ and $c_1,c''_1\in[a_1,a_1+n_1)$ be such that $p_0[c_0]=z_0$, $p_0[c''_0]=z''_0$, $p_1[c_1]=z_1$, and $p_1[c''_1]=z''_1$.
This and $z_0z_1\prec z''_0z''_1$ imply $c_0<c''_0$ and $c_1<c''_1$.
This and the assumption that $s_0\leq c_0$ and $s_1\leq c_1$ (which is the invariant before the update) imply $s_0<c''_0$ and $s_1<c''_1$.
This already gives one inequality of the invariant after the update, namely, $s_{1-u}\leq c''_{1-u}$.
It remains to prove $v_u\leq c''_u$, which is the other inequality of the invariant after the update.
Since $q''_u\notin\RHP(q_0,q_1)$ and $q''_{1-u}=q_{1-u}$, we have $\RHP(q_0,q_1)\cap q''_0q''_1=\{q''_{1-u}\}$.
This and $P_u[q_u,q'_u]\subset\RHP(q_0,q_1)$ imply $P_u[q_u,q'_u]\cap q''_0q''_1=\emptyset$.
Since $P_u[q'_u,q''_u]=q'_uq''_u$ and $q'_u\notin q''_0q''_1$, we have $P_u[q'_u,q''_u]\cap q''_0q''_1=\{q''_u\}$.
Thus $P_u[q_u,q''_u]\cap q''_0q''_1=\{q''_u\}$.
This and the fact that $p_u[c''_u]=z''_u\in P_u\cap q''_0q''_1$ imply $c''_u\notin[s_u,v_u)$.
This and $s_u<c''_u$ give the requested inequality $v_u\leq c''_u$.
We conclude that the invariant is preserved at the considered update to $s_u$.
\end{proof}

\section{Concluding remarks}
\label{sec:conclusion}

So far, we were assuming that the combined set $\smash[t]{\hat P}$ of corners of $P_0$ and $P_1$ contains no triple of collinear points, which guarantees that expressions of the form $\dets(a,b,c)$ with $a,b,c\in\smash[t]{\hat P}$ evaluated in line~\ref{line1:side} of Algorithm~\ref{ALG1} and in lines \ref{line2:side} and~\ref{line2:triangle} of Algorithm~\ref{ALG2} are non-zero.
Now, we adapt the algorithms to handle the case that $\smash[t]{\hat P}$ may contain triples of collinear points.
Consider the following family of transformations:
\begin{equation*}
\phi_\epsilon\colon\setR^2\ni(x,y)\mapsto\bigl(x+\epsilon y,y+\epsilon(x+\epsilon y)^2\bigr)\in\setR^2.
\end{equation*}
For any distinct points $a,b,c\in\setR^2$, the expression $\dets(\phi_\epsilon(a),\phi_\epsilon(b),\phi_\epsilon(c))$ is a continuous function of $\epsilon$ that attains value zero for finitely many arguments $\epsilon$.
Therefore, there is $\epsilon_0>0$ such that the sign of $\dets(\phi_\epsilon(a),\phi_\epsilon(b),\phi_\epsilon(c))$ is equal to the same constant $\sigma(a,b,c)\in\{+1,-1\}$ for all $\epsilon\in(0,\epsilon_0)$.
The value of $\sigma(a,b,c)$ can be easily computed from the coordinates of $a$, $b$, and $c$ by treating $\epsilon$ as a symbolic positive infinitesimal.
We can modify Algorithms \ref{ALG1} and~\ref{ALG2} to use $\sigma(a,b,c)$ instead of $\dets(a,b,c)$ for the tests in the aforementioned lines, so that $c\notin\HHH_k(a,b)$ means $\sigma(a,b,c)=\alpha_k$, and $z\in\Delta(a,b,c)$ means $\sigma(z,a,b)=\sigma(z,b,c)=\sigma(z,c,a)$.
Conceptually, this is the same as invoking the algorithms on the polygons $P_0$ and $P_1$ transformed by $\phi_\epsilon$, where $\epsilon$ is small enough so that $\sigma(a,b,c)=\sgn\dets(\phi_\epsilon(a),\phi_\epsilon(b),\phi_\epsilon(c))$ for all triples of distinct points $a,b,c\in\smash[t]{\hat P}$.
Therefore, if either algorithm claims to find a solution, the fact that it is correct on the transformed input implies that it is correct on the original input.
The same reasoning (and the same modification of the two algorithms) can be applied with the following family of transformations $\psi_\epsilon$ instead of $\phi_\epsilon$:
\begin{equation*}
\psi_\epsilon\colon\setR^2\ni(x,y)\mapsto\bigl(x+\epsilon y,y-\epsilon(x+\epsilon y)^2\bigr)\in\setR^2.
\end{equation*}
Any straight line is transformed by $\phi_\epsilon$ and $\psi_\epsilon$ (with $\epsilon$ small enough) into two curve lines bend in the opposite directions.
This property is important when we want to find ``degenerate'' common tangents.
Namely, if the convex hulls of $P_0$ and $P_1$ touch, then one of $\phi_\epsilon$ and $\psi_\epsilon$ makes them overlap properly while the other makes them disjoint; therefore, one modification of Algorithm \ref{ALG1} or~\ref{ALG2} finds the ``degenerate'' separating common tangent while the other does not.
Similarly, if the convex hulls of $P_0$ and $P_1$ are nested and their boundaries touch, then one of $\phi_\epsilon$ and $\psi_\epsilon$ makes them overlap properly while the other moves the smaller convex hull to the interior of the larger; therefore, one modification of Algorithm~\ref{ALG2} finds the ``degenerate'' outer common tangent while the other does not.
We recognize that there is no solution when both modifications report no solution.

It remains open whether an outer common tangent of two polygons that are not disjoint can be found in linear time and constant workspace.
Another natural question is whether the diameter of the convex hull of a simple polygon can be computed in linear time and constant workspace.

\bibliographystyle{ACM-Reference-Format}
\bibliography{tangents}

\end{document}